\documentclass{jfm}
\usepackage{graphicx}
\usepackage{epstopdf, epsfig}
\usepackage{natbib}
\usepackage[dvipsnames]{xcolor}
\usepackage{amssymb}
\usepackage{amsmath}
\usepackage{bm} 
\usepackage{latexsym}
\usepackage{lineno}

\newcommand\Reytau{\mathrm{Re}_{\tau}}
\newcommand\dd{\mathrm{d}}

\shorttitle{effect of the opposite wall in ducted turbulent flows}

\shortauthor{P. A. Monkewitz}

\title{The late start of the mean velocity overlap layer at $\mathbf{y^+=\mathcal{O}(10^3)}$
- A generic feature of ducted turbulent flows}

\author{Peter A. Monkewitz
  \corresp{\email{peter.monkewitz@epfl.ch}}}

\affiliation{\'Ecole Polytechnique F\'ed\'erale de Lausanne (EPFL), CH-1015, Lausanne, Switzerland}

\begin{document}

\maketitle   

\begin{abstract}
One of the key observations in the Princeton Superpipe was the late start of the logarithmic mean velocity overlap layer at a wall distance of the order of $10^3$ inner units. Between $y^+\approx 150$, the start of the overlap layer in zero pressure gradient turbulent boundary layers, and $y^+\approx 10^3$, the Superpipe profile is modeled equally well by a power law or a log-law with a larger slope than in the overlap layer. In this paper it is shown that the mean velocity profile in turbulent plane channel flow exhibits analogous characteristics, namely a sudden decrease of logarithmic slope (increase of $\kappa$) at a $y^+_{\mathrm{break}}\approx 600$, which marks the start of the actual overlap layer. This demonstration results from the first construction of the complete inner and outer asymptotic expansions up to order $\mathcal{O}(\Reytau)^{-1}$ from mean velocity profiles of direct numerical simulations (DNS) at moderate Reynolds numbers. A preliminary analysis of a Couette flow DNS, on the other hand, yields an increase of logarithmic slope (decrease of $\kappa$) at a $y^+_{\mathrm{break}}\approx 400$. The correlation between the sign of the slope change and the flow symmetry motivates the hypothesis that the breakpoint between the possibly universal short inner logarithmic region and the actual overlap log-law corresponds to the penetration depth of large-scale turbulent structures originating from the opposite wall.
\end{abstract}

\section{\label{sec1}Aim of the study and some comments on the asymptotic analysis of mean velocity profiles}

The present study aims to unify the modeling of mean velocity profiles in terms of asymptotic expansions for the three ``canonical'' turbulent parallel flows: pipe, plane channel and plane Couette flows, i.e. flows which are in the mean homogeneous in both the stream-wise and azimuthal/span-wise directions. All three flows are characterized by a single ``outer'' or global length scale $\breve{L}$, the pipe radius or channel half-width, and a constant wall shear stress $\breve{\tau}_w$, where $\breve{\cdot}$ identifies dimensional quantities throughout the paper. In the following, the classical two-layer description is adopted with the standard ``inner'' or viscous length scale $\breve{\ell} \equiv (\breve{\nu}/\breve{u}_\tau)$, where $\breve{u}_\tau \equiv (\breve{\tau}_w/\breve{\rho})^{1/2}$, $\breve{\rho}$ and $\breve{\nu}$ are the friction velocity, density and dynamic viscosity, respectively. The relevant Reynolds number is the ``friction Reynolds number'' $\Reytau\equiv \breve{L}/\breve{\ell}$.

To date, only the Princeton Superpipe mean velocity profiles \citep{ZS97,ZS98,McK_thesis,MLJMS04}, reviewed in section \ref{sec21}, were acquired at high enough Reynolds numbers to reveal the detailed structure of the overlap layer, which links the wall layer to the core. The characteristic feature of this layer for friction Reynolds numbers $\Reytau$ beyond about 30'000 - a distinct reduction of (logarithmic) slope at a wall distance well beyond the start of the log-law in zero pressure gradient turbulent boundary layers (abbreviated ZPG TBL's) - is reviewed in section \ref{sec21}. In section \ref{sec22}, the hypothesis is advanced that this feature is related to ``eddies'' emanating from the opposite wall. The outline of the more technical continuation of the paper is postponed to section \ref{sec23}.

Since the interpretation of channel and Couette flow experiments is complicated by the finite span-wise extent of facilities \citep[see e.g.][]{Vinuesa2018} as well as by relatively low Reynolds numbers, attention is turned to DNS. As the Reynolds numbers for the available DNS are relatively low, the contamination of the overlap mean velocity profile by both its inner and outer parts remains a problem. This ``Reynolds number handicap'' of DNS is overcome by constructing, within the framework of matched asymptotic expansions (abbreviated MAE), higher order inner and outer expansions (in practice two-term expansions) of mean velocity in order to reveal the infinite Reynolds number limit at the leading order. However, before getting into the technical details of constructing such expansions from DNS, it is useful to review some basic principles of MAE \citep[see for instance the excellent monograph of][]{KC85} and their application to wall turbulence, reviewed, for instance, by \citet{Panton2005}.

Within the framework of MAE, the non-dimensional mean velocity $U^+ \equiv \breve{U}/\breve{u}_\tau$ at large $\Reytau$ is modelled by inner and outer asymptotic expansions $U^+_{\mathrm{in}}(y^+)=\sum \phi_n(\Reytau) f_n(y^+)$ and $U^+_{\mathrm{out}}(Y)=\sum \Phi_n(\Reytau) F_n(Y)$, where $y^+ \equiv \breve{y}/\breve{\ell}$ and $Y\equiv \breve{y}/\breve{L}=y^+/\Reytau$ are the inner-scaled and outer-scaled non-dimensional wall-normal coordinates, while $\phi_n(\Reytau)$ and $\Phi_n(\Reytau)$ are suitable gauge functions. These inner and outer expansions for $U^+$ have to be matched in an ``overlap'' layer, where $(y^+Y)$ is of order unity.

This overlap layer is however NOT a third layer on the same footing as inner and outer layers, but the ``intersection'' or the common part of the inner and outer expansions. As its name suggests, it only contains terms that are \textbf{common to both inner and outer expansions} and their number depends therefore on how many terms are retained in the two expansions to be matched. This precise definition allows to construct the additive composite profile, which is the sum of inner and outer expansions minus the common part, as the latter is counted twice in the sum.

An important corollary to this statement is that the common part contains no new physics, unless it is introduced by an additional reasoning. The classical example in the present context is the postulate of asymptotic independence of inner and outer scales by \citet{ICTAM05} \citep[see also the early formulation by][]{vonKarman30}, from which it follows that, in the overlap layer, $\,y^+(\dd U^+/\dd y^+) = Y(\dd U^+/\dd Y)$ can only be a constant $\kappa^{-1}$. This physical argument yields directly the functional form of the leading order common part, the log-law
\begin{equation}
U^+_{\mathrm{cp,0}} = \kappa^{-1}\,\ln(y^+) + B = \kappa^{-1}\,\ln(Y) + B + \kappa^{-1}\,\ln(\Reytau)
\label{loglaw}
\end{equation}

A first remark concerns the log-law in outer variables, which contains both order $\mathcal{O}(1)$ terms and a $\ln(\Reytau)$. Therefore it must be regarded as of ``block order'' unity, where the block order, introduced by \citet{CL73}, regroups all the terms of order $\epsilon^n \ln^m(\epsilon)$ with different $m$'s into a single block order $\epsilon^n$. This is also seen in the ``law of the wake'' of \citet{Coles56}, $U^+_{\mathrm{out,0}} = \kappa^{-1}\ln(\Reytau) + \kappa^{-1}\ln(Y) + B + 2\pi\,\kappa^{-1} f(Y)$, which is one way of writing the leading term of the outer expansion.

The reported value of $\kappa$ in equation (\ref{loglaw}) has varied considerably between different flows and over time, from the 0.38 originally estimated by \citet{vonKarman30} to the ``popular'' value of 0.41 \citep[see for instance][section 7.3.3]{Pope2000} to 0.436 in the Superpipe \citep{ZS98} and the CICLoPE pipe \citep{FioriniPhD,NagibCICLoPE} \citep[see for instance the extensive discussion in][]{Metal2010}.

The diversity of $\kappa$ values should however not come as a surprise, since the Millikan argument does in no way preclude the dependence of $\kappa$ and $B$ in equation (\ref{loglaw}) on control parameters, such as for instance the pressure gradient parameter $\beta \equiv -\,\breve{L} \,\breve{p}_x\, (\breve{\tau}_w)^{-1}$ ($\beta$ = 0, 1 and 2 for Couette, channel and pipe flow, respectively), geometry, etc. \citep[see e.g.][]{NagibChauhan2008}. Furthermore, its value in different flows is still difficult to pin down because the $\kappa$'s extracted from high Reynolds number experiments come with a significant uncertainty \citep[see e.g.][]{Bailey14}, while the $\Reytau$ of high quality DNS are still too low to produce clean log-laws.

Possibly because of this uncertainty, $\kappa$ has retained an aura of fundamental constant, the K\'arm\'an ``constant'', and prompted considerable attention to higher order terms in the overlap region by, among others, \citet{Yajnik70}, \citet{AfzalY73}, \citet{Jimenez07} and most recently by \citet[see appendix \ref{App2} for a critical appraisal]{Luchini17}.

The main points on asymptotic matching and common parts may be summarized as follows :
\begin{enumerate}
\item The logarithm in the \textbf{leading-order} overlap profile for $U^+$ is a consequence of the postulated asymptotic independence of inner and outer scales and hence of physical origin.
\item Higher order terms in the common part depend entirely on which terms are included in the inner and outer expansions, as they must be contained in \textbf{both} the limits $y^+\gg 1$  of the inner and $Y\ll 1$ of the outer expansion.
\item It follows directly from equation \ref{loglaw}, that the $\kappa$'s determined from the leading order overlap profile and the leading order centerline velocity $U^+_{\mathrm{CL,0}}(\Reytau) = \kappa^{-1}\ln(\Reytau) + C$ \textbf{must be identical} !
\end{enumerate}

\section{\label{sec2}Reconciling the mean velocity profile
of the Princeton Superpipe with plane channel and Couette profiles}
\subsection{\label{sec21}The principal characteristics of the mean velocity overlap profile in the Superpipe}

Until the Princeton Superpipe experiment of \citet{ZS97,ZS98}, the ``standard model'' of the mean velocity profile in wall-bounded turbulent flows consisted of inner and outer profiles, monotonically connected by the logarithmic overlap profile extending from $y^+\approx 150$ to $Y\approx 0.2$, except for a small overshoot centered around $y^+\approx 30$ \citep[see][and appendix \ref{sec:App1}]{NagibChauhan2008}. It goes without saying that the extent of the overlap region depends on how much deviation from the pure log-law is tolerated.

\begin{figure}
\center
\includegraphics[width=0.7\textwidth]{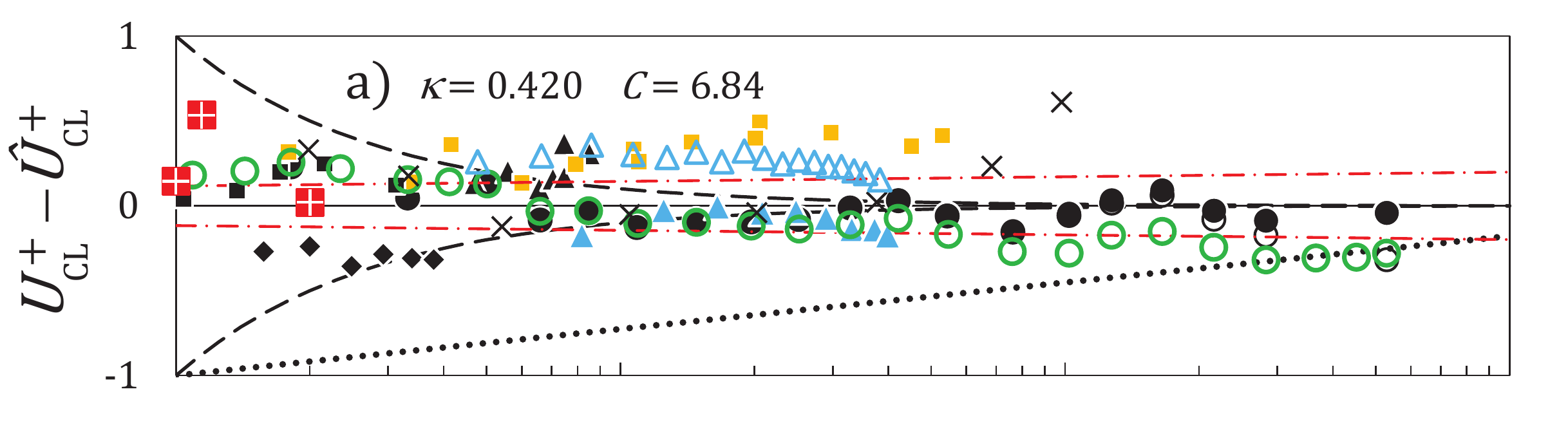}
\includegraphics[width=0.7\textwidth]{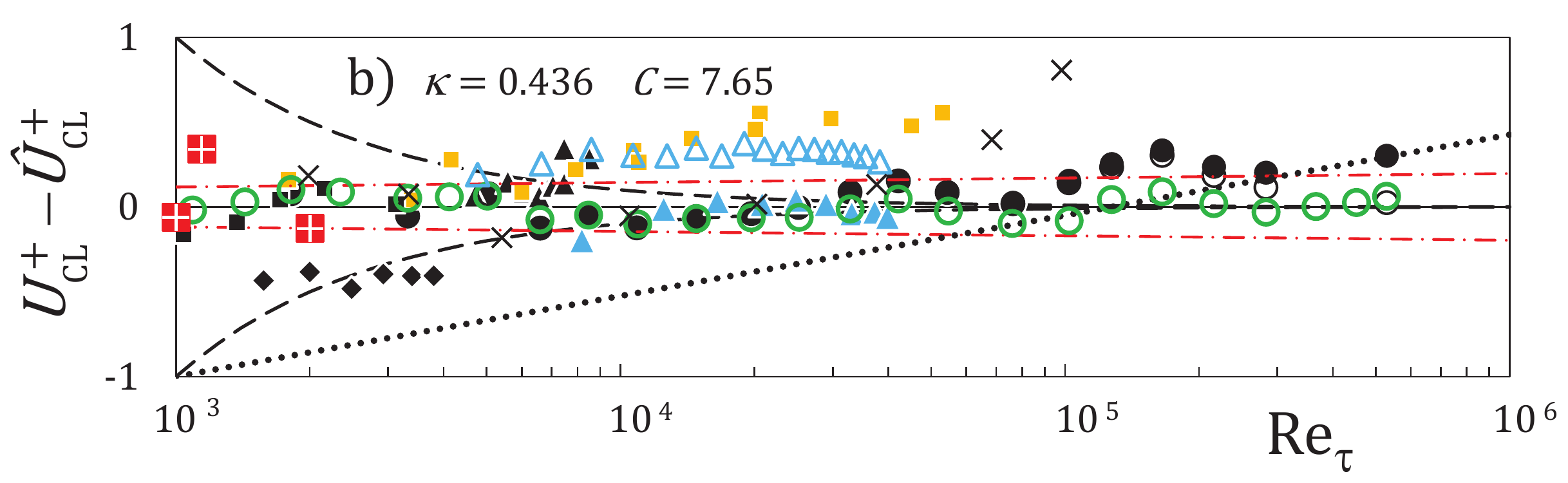}
\caption{\label{Fig:PCL} Pipe centerline velocities minus $\kappa^{-1}\ln(\Reytau)+C$ versus $\Reytau$ for (a) $\kappa=.42$ , $C=6.84$ and (b) $\kappa=.436$ , $C=7.65$.  $\bullet$, Superpipe data corrected according to McKeon; $\circ$, same data without roughness correction; \textcolor{green}{$\circ$}, Superpipe data of \citet{ZS97} with same roughness correction; $\times$, Superpipe NSTAP data of \citet{Hultetal12}; $\blacklozenge$, \citet{PA77}; $\blacktriangle$, \citet{zanoun2007}; $\blacksquare$, \citet{Monty_thesis}; \textcolor{blue}{$\triangle \triangle \triangle$}, CICLoPE data of \citet{FioriniPhD}; \textcolor{blue}{$\blacktriangle \blacktriangle \blacktriangle$}, new CICLoPE data of \citet{NagibAPS,NagibCICLoPE}; \textcolor{Yellow}{$\blacksquare \blacksquare \blacksquare$}, fig. 6 of \citet{Furuichi18}; \textcolor{red}{$\blacksquare$}, the three DNS of \citet{ElKhoury2013} ($\Reytau=999$), \citet{WuMoin08} ($\Reytau=1142$) and \citet{ChinPipe14} ($\Reytau=2003$). \textcolor{red}{$\cdot - \cdot$}, $\pm 0.5\%$ of reference $\hat{U}^+_{\mathrm{CL}}$; - - -,  $\pm 10^3/\Reytau$; $\cdot\cdot\cdot$, slope corresponding to $\kappa =0.40$.}
\end{figure}

The challenge to this ``standard model'' by the Princeton Superpipe experiment has been twofold:
\begin{enumerate}
\item The originally reported $\kappa$ of 0.436, as well as the revised value of 0.421, obtained by \cite{MLJMS04} with smaller Pitot probes and a different correction scheme have attracted a great deal of scepticism. Based on the extensive collection of centerline velocities $U^+_{\mathrm{CL}}$ minus the fit $\kappa^{-1}\ln(\Reytau)+C$ in figure \ref{Fig:PCL} for the two different $\kappa$'s and corresponding $C$'s, the pipe overlap $\kappa$ can be placed in the bracket $[0.42, 0.44]$. With the present data, it is not possible to pinpoint it more precisely \citep[see e.g.][]{Bailey14}. It is however clear that the Superpipe $\kappa$ and the preliminary values from the CICLoPE facility \citep{FioriniPhD,NagibCICLoPE}
    are different from the $\kappa$ in ZPG TBL's, which has converged to around 0.384 \citep[see for instance][]{MCN07,Metal2010}. However, in light of the comments on the Millikan matching argument in section \ref{sec1}, these differences should not come as a surprise as they do not violate any basic principles.

\item More importantly, the overlap log-law was found to start only beyond $y^+ \gtrapprox 500$, much further from the wall than in the ZPG TBL, where a clean log-law is observed for $y^+\gtrapprox 150$ \citep[see for instance][]{MCN07,Metal2010}. Surprisingly, this feature of the Superpipe profiles has gone largely uncommented and certainly unexplained.

    Originally, both \citet{ZS98} and \cite{MLJMS04} have fitted $U^+$ in the interval $150\lessapprox y^+\lessapprox 500$ with power laws, but \citet{McK_thesis} noted that a logarithm with slope $1/0.385$ also ``fits quite well''. Consistent with this observation, the hypothesis (\ref{hyp}) and the analysis of section \ref{sec33}, both the near-wall and the overlap region will be modeled by logarithmic laws with log-slopes of $(1/\kappa_\mathrm{M})$ and $(1/\kappa)$, respectively, and a rather sharp transition between the two at a $y^+_{\mathrm{break}}$ of around 500 (note that the ``M'' in $\kappa_{\mathrm{M}}$ indicates that it is the kappa used to generate the inner Musker profile of appendix \ref{sec:App1}).

\item Based on velocity measurements with miniature ``NSTAP'' hotwires in the Princeton Superpipe, \citet{MMHS13} have put the breakpoint $y^+_{\mathrm{break}}$, i.e. the start of the logarithmic overlap region, at $y^+_{\mathrm{break}} \approx 3\,\Reytau^{1/2}$. On the other hand, \citet{Monk17} found that, based on the original Pitot measurements, the slope change correlated better with a fixed $y^+_{\mathrm{break}} \approx 500$. At the $\Reytau$ considered, these two scalings for the start of the overlap log-law are numerically similar and within the uncertainty of the breakpoint location. However, the scaling on the intermediate variable $y^+ \Reytau^{-1/2}$ poses a problem: if the inner profile for $y^+<y^+_{\mathrm{break}}$ is a function of $y^+$ alone, as observed, the additive log-law constant $B$ can no longer be constant, but increases with $\ln(\Reytau)$. Furthermore, the comparison of the two scalings over the full Superpipe Reynolds number range by \citet{Monkarxiv19} clearly favors a constant $y^+_{\mathrm{break}}$, which is therefore adopted in the following.\newline
\end{enumerate}

\begin{figure}
\center
\includegraphics[width=0.8\textwidth]{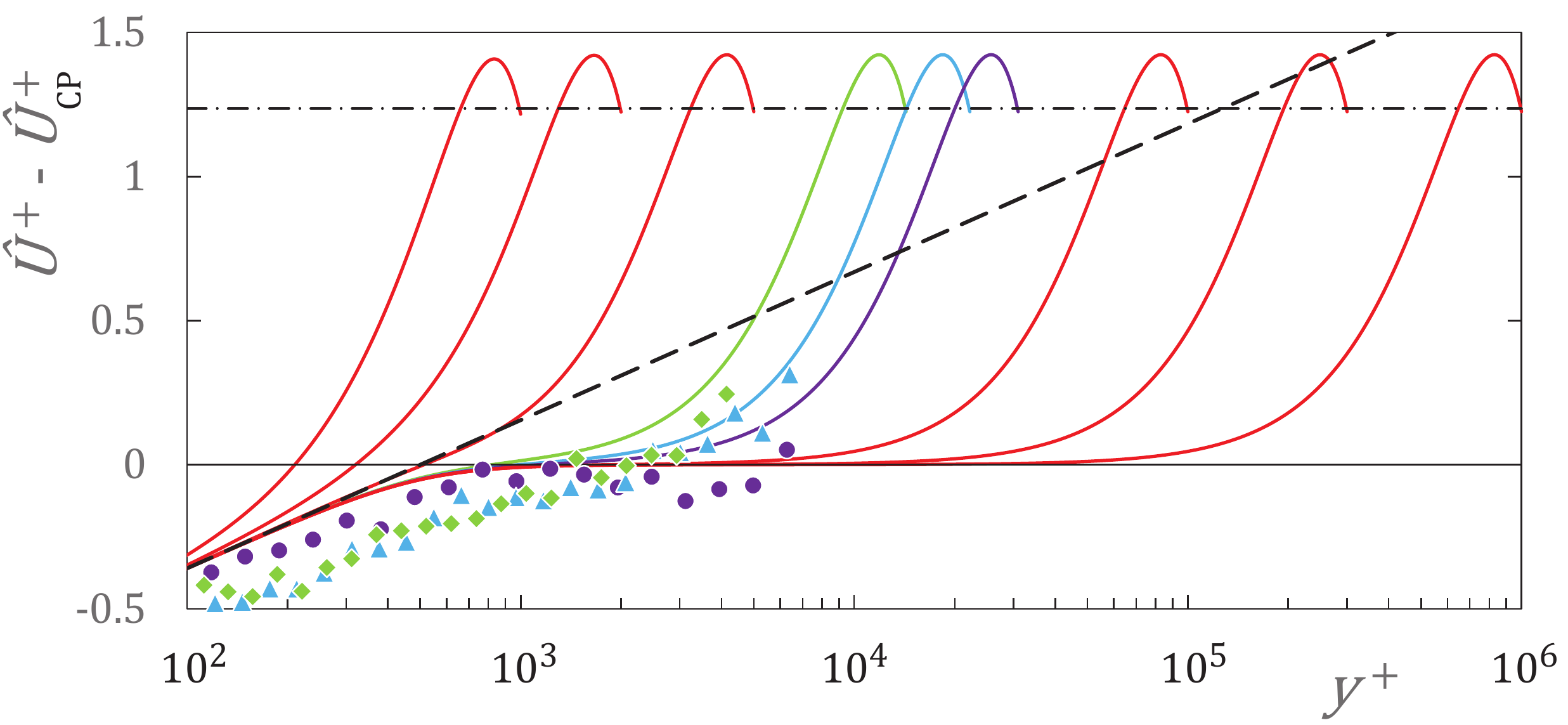}
\caption{(color online)\ Model pipe flow profiles $\hat{U}^+ - \hat{U}^+_{\mathrm{CP}}$ from \citet{Monk17} with $\hat{U}^+_{\mathrm{CP}} = (1/0.42) \ln(y^+) + 5.604$. \textcolor{Red}{---}, $\Reytau = 1, 2, 5, 100, 300, 1000 \times 10^3$. \textcolor{LimeGreen}{---, $\blacklozenge \blacklozenge$}, \textcolor{SkyBlue}{---, $\blacktriangle \blacktriangle$}, \textcolor{Violet}{---, $\blacksquare \blacksquare$}, model profiles and hotwire data of \citet{FioriniPhD} for $\Reytau = 14.3, 22.2, 31.0 \times 10^3$. $- \cdot -$ , $\hat{U}^+_{\mathrm{CL}} - \hat{U}^+_{\mathrm{CP}} = 1.24$; - - - , asymptote $[(1/0.384)-(1/0.42)]\,\ln(y^+/500)$ for the deviation of the inner logarithmic part of the profile from $\hat{U}^+_{\mathrm{CP}}$. }
\label{figpipeout}
\end{figure}

The main Superpipe findings, detailed above, are illustrated in figure \ref{figpipeout} by the model profiles of \cite{Monk17} with the original parameters indicated in the figure caption. Note that in this figure, and in the rest of the paper, profile fits are identified by hats $\,\hat{\cdot}\,$, while profiles derived from experimental or DNS data have no hat. Also included in the figure are the first partial velocity profiles obtained by \citet{FioriniPhD} in the CICLoPE pipe with traditional 1 and 1.1mm hotwires. These hotwire data are consistent with a sudden increase of $\kappa$ by 0.03-0.04, seen between $y^+$ of 500 and 1000 for $\Reytau$  beyond about $30'000$, but one will have to wait for an upgraded CICLoPE instrumentation to obtain more precise experimental values for $\kappa_\mathrm{M}$, $y^+_{\mathrm{break}}$ and $\kappa$ (see section \ref{sec42} for the prospect of using DNS).

The Superpipe results described above were met with scepticism, to say the least, and the interrogations were numerous: The question of corrections for wall roughness was brought up by \cite{perry2001} and finally resolved by \cite{Allen05}. The diverse Pitot probe corrections were questioned and prompted a vast investigation by an international collaboration \citep{Pitot13}. Finally, the effect of Pitot tube positioning errors was considered by \cite{VinDN16}. In the end, the Superpipe results have withstood all these additional investigations, and so one has to ask whether the mean velocity profile in other ducted parallel flows, in particular plane channel and Couette flow, will also exhibit the Superpipe features of figure \ref{figpipeout} if pushed to higher Reynolds numbers. This author cannot conceive of any reason for this not to be the case, and so the Superpipe mean velocity structure is expected to also emerge at higher $\Reytau$ in plane channel and Couette flows. Before demonstrating that the Superpipe profile features are also found in plane channel and Couette flows, it is helpful to also think of a coherent explanation for the differences between the logarithmic regions in ZPG TBL's and ducted parallel flows. Such an explanation is proposed in the following section \ref{sec22}.

\subsection{\label{sec22}Hypothesis on the effect of the opposite wall}

The following explanation is proposed for both the late start of the overlap log-law and the flow dependence of the overlap $\kappa$ in simple ducted parallel flows, illustrated in figure \ref{figpipeout} for the pipe:

\begin{equation}\label{hyp}
\mathrm{\textbf{HYPOTHESIS}}
\end{equation}
\nopagebreak
\textbf{The breakpoint $\mathbf{y^+_{\mathrm{break}}}$, separating the short logarithmic region with slope $\mathbf{(1/\kappa_{\mathrm{M}})}$ between $\mathbf{y^+\approx 150}$ and $\mathbf{y^+_{\mathrm{break}}}$ and the true overlap log-law with K\'arm\'an parameter $\mathbf{\kappa}$, corresponds to the penetration depth of large scale turbulent structures originating from the opposite wall.}\newline

\begin{figure}
\center
\hfill\includegraphics[width=0.35\textwidth]{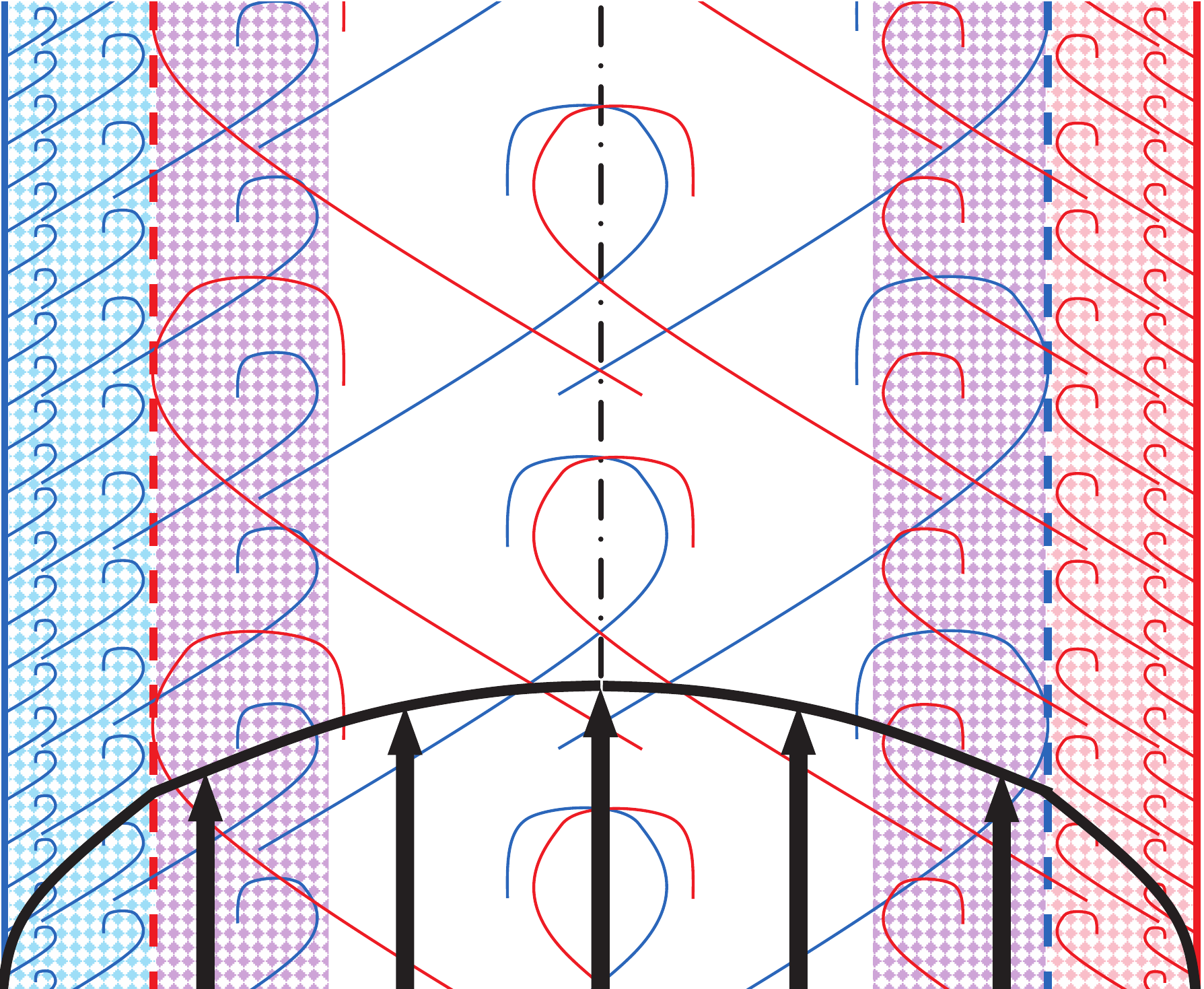}%
\hfill\includegraphics[width=0.35\textwidth]{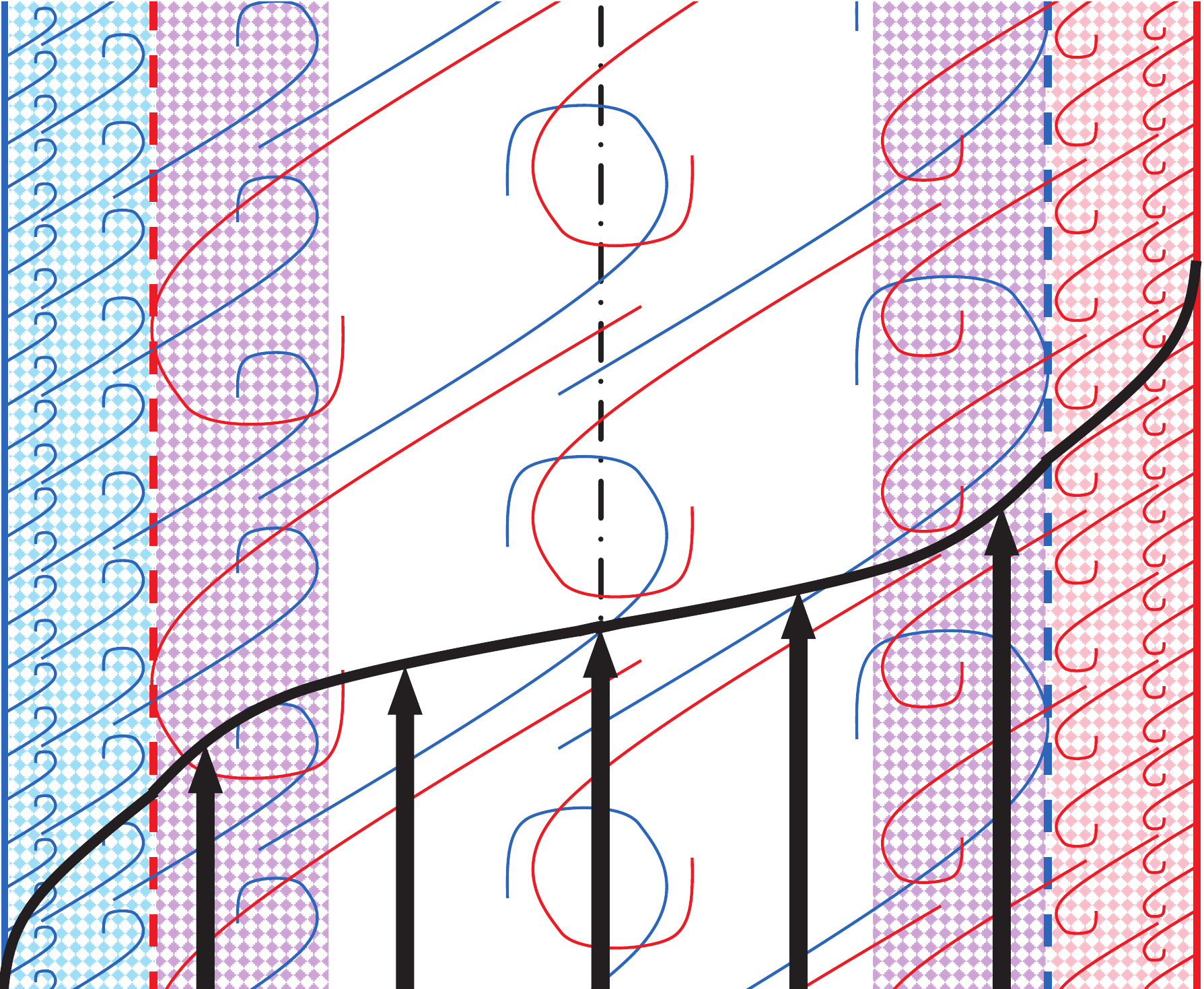}\hspace*{\fill}
\caption{Cartoons illustrating the hypothesis \ref{hyp}. Red and blue shaded areas: wall layers not affected by the opposite wall; violet shade: overlap layers affected by ``eddies'' originating from the opposite wall. Left cartoon: overlap logarithmic slope reduced relative to the wall layers for pipe and channel flows; Right cartoon: overlap logarithmic slope increased relative to the wall layers for Couette flow.}
\label{hypfig}
\end{figure}

\noindent This hypothesis (\ref{hyp}) is visualized by the cartoon of figure \ref{hypfig} and has two testable consequences :
\begin{enumerate}
\item In pipe and channel flows the vorticity emanating from the opposite wall reduces $\dd U^+/\dd y^+$ for $y^+ > y^+_{\mathrm{break}}$ and hence $\kappa > \kappa_{\mathrm{M}}$. Conversely, in Couette flow this vorticity must increase the mean shear outside of $y^+_{\mathrm{break}}$, leading to $\kappa < \kappa_{\mathrm{M}}$.
\item At sufficiently high $\Reytau$, the short logarithmic layer with slope $(1/\kappa_{\mathrm{M}})$ is not influenced by geometry and the inner layer $0 \leq y^+ \leq  y^+_{\mathrm{break}}$ may therefore be universal, at least for the truly parallel flows considered here.\newline
\end{enumerate}

\subsection{\label{sec23}Outline of sections 3 to 5}

In the next section \ref{sec3}, a number of high quality profiles for plane channel flow, up to $\Reytau=5186$ \citep{LM14}, are used to construct, for the first time, the complete inner and outer asymptotic expansions of $U^+$ up to terms of order $\mathcal{O}(\Reytau)^{-1}$. The final result for the leading order inner profile $U^+_{\mathrm{in, 0}}$ in section \ref{sec33} corroborates the hypothesis (\ref{hyp}) by revealing a clean break point $y^+_{\mathrm{break}} \approxeq 600$, where the logarithmic slope of $U^+$ decreases abruptly from (1/0.398) to (1/0.42).

An analogous reconstruction of 2-term asymptotic expansions from available Couette DNS, on the other hand, has not been feasible. Limited to the leading order inner and outer asymptotic expansions, it is nevertheless possible to demonstrate in section \ref{sec41}, that the Couette DNS of \citet{krah2018} for $\Reytau=1026$ does show the steepening of $U^+$ at $y^+_{\mathrm{break}}$, corresponding to $\kappa=0.367 < \kappa_{\mathrm{M}}=0.40$, in conformity with the hypothesis (\ref{hyp}).

A brief review of three pipe DNS profiles in section \ref{sec42} finally reveals, that the differences between available profiles are too large to attempt an analysis analogous to the one for the channel.

The paper closes with a recap of selected findings and some open questions in section \ref{sec5}.

\section{\label{sec3}Higher order asymptotic expansions of $U^+$ for the plane channel}

\subsection{\label{sec31}Methodology for extracting asymptotic expansions from DNS}

The objective is to obtain, for the plane channel, the inner and outer asymptotic expansions of the mean velocity $U^+$ up to and including the block order $\mathcal{O}(\Reytau^{-1})$ \citep[see][and section \ref{sec1} for the concept of block order]{CL73}
\begin{align}
U^+_{\mathrm{in}}(y^+) &= U^+_{\mathrm{in, 0}}(y^+) + \Reytau^{-1}\, U^+_{\mathrm{in, 1}}(y^+) + \mathcal{O}(\Reytau^{-2}) \label{Uin} \quad \mathrm{and} \\
U^+_{\mathrm{out}}(Y) &= U^+_{\mathrm{out, 0}}(Y) + \Reytau^{-1}\, U^+_{\mathrm{out, 1}}(Y) + \mathcal{O}(\Reytau^{-2}) \quad ,
\label{Uout}
\end{align}
together with the common part $U^+_{\mathrm{cp}}$, which can be expressed in terms of $y^+$, $Y$, or the intermediate variable $\eta=y^+\Reytau^{-1/2}=Y\,\Reytau^{1/2}$. Identifying the composite expansion $U^+_{\mathrm{comp}}=U^+_{\mathrm{in}}+U^+_{\mathrm{out}}-U^+_{\mathrm{cp}}$ with $U^+_{\mathrm{DNS}}$, the first two orders in the expansions (\ref{Uin}) and (\ref{Uout}) are successively determined for the first time and fitted by suitable functions.

Rather counter-intuitively, the determination of the inner and outer expansions is best started with the order $\mathcal{O}(\Reytau)^{-1}$ terms. Between the wall and the overlap region, the deviation of the inner velocity (equ. \ref{Uin}) from the total velocity, taken to be $U^+_{\mathrm{DNS}}(y^+)$, is of the order of $\mid U^+_{\mathrm{out}}(Y)-U^+_{\mathrm{cp}}\mid$. Hence, assuming that the asymptotic expansion (\ref{Uin}) converges rapidly (an assumption justified a posteriori), one obtains a good estimate of $U^+_{\mathrm{in, 1}}(y^+)$ between the wall and the overlap layer by taking differences of two total velocity profiles at equal $y^+$'s (obtained by 3-point quadratic interpolation of the original DNS data) and different $\Reytau$'s :
\begin{align}
&\left[U^{+}_{\mathrm{DNS}}\left(y^+;\mathrm{Re}_{\tau, 1}\right) - U^{+}_{\mathrm{DNS}}\left(y^+;\mathrm{Re}_{\tau, 2}\right)\right] \left[\mathrm{Re}^{-1}_{\tau, 1} - \mathrm{Re}^{-1}_{\tau, 2}\right]^{-1} = \nonumber \\ &\quad U^+_{\mathrm{in}, 1}(y^+) + \mathcal{O}(\Reytau^{-1}\,; \mid U^+_{\mathrm{out}}(Y) - U^+_{\mathrm{cp}}\mid)
\label{diffin}
\end{align}

Similarly, between the overlap region and the centerline, the outer velocity (equ. \ref{Uout}) is equal to the total velocity $U^+_{\mathrm{DNS}}(Y)$, with an error of order $\mid U^+_{\mathrm{in}}(y^+)-U^+_{\mathrm{cp}}\mid$. However, obtaining the first order term $U^+_{\mathrm{out, 1}}(Y)$ is a bit trickier, because the leading block order of the outer expansion (\ref{Uout}) is of the well-known form
\begin{equation}
U^+_{\mathrm{out, 0}}(Y) = (1/\kappa) \ln(\Reytau) + F(Y) .
\label{Uout0}
\end{equation}

Two strategies to determine $U^+_{\mathrm{out, 1}}(Y)$ are pursued in the next section \ref{sec32} :\newline
\begin{enumerate}
\item The first is to assume $\kappa$ in equation (\ref{Uout0}) and to use the analogue of equation (\ref{diffin}) to determine $U^+_{\mathrm{out, 1}}(Y; \kappa)$. The ``true'' $U^+_{\mathrm{out, 1}}(Y)$ is then obtained by iterating on $\kappa$ until the best collapse of $U^+_{\mathrm{out, 1}}(Y)$ is obtained from different profile pairs.
\item The second, completely parameter-free strategy is to use a third DNS profile at a different Reynolds number to eliminate the $(1/\kappa) \ln(\Reytau)$ term from the two DNS profiles at $\mathrm{Re}_{\tau, 1}$ and $\mathrm{Re}_{\tau, 2}$, before proceeding analogous to equation (\ref{diffin}).\newline
\end{enumerate}

The strategies outlined above to educe higher order terms from DNS, are fundamentally different from the attempts to determine higher order terms in the \textbf{overlap} region discussed in section \ref{sec1}. Here, $U^+_{\mathrm{in, 1}}(y^+)$ and $U^+_{\mathrm{out, 1}}(Y)$ are determined from profiles in the inner wall-region and the outer region near the centerline, respectively. Their proper matching in the overlap region only serves as an a posteriori verification.

The primary difficulty in developing these asymptotic expansions is the uncertainty of the DNS, which must be sufficiently smaller than $U^+/\Reytau$ in order to extract the $\mathcal{O}(\Reytau)^{-1}$ terms with any kind of confidence. At first thought, one might wish for higher DNS Reynolds numbers in order to obtain a better separation of inner and outer scale and hence a clean(er) overlap log-law. However, if the uncertainty of the DNS does not diminish at least as $1/\Reytau$, nothing is gained for the determination of higher order terms in the asymptotic expansion. In other words, it appears more important to improve the fidelity of DNS than to keep increasing the Reynolds number. Finally, the dependence of the $U^+$ profiles on additional parameters, such as the computational box size, has to be much weaker than its dependence on $\Reytau$.

Four DNS profiles listed in table \ref{TableDNS} have been found suitable for the construction of the asymptotic expansions (\ref{Uin}, \ref{Uout}) and will in the following be referred to by their profile number in the table.

\begin{table}
\center
\caption{Channel DNS profiles used}
\begin{tabular}{l l l l l}
\hline
Profile $\quad$ & \#1 & \#2 & \#3 & \#4 \\
$\Reytau$ & 5186 &2004 & 1000 & 934 \\
Ref. & Lee \& Moser & Hoyas \& Jim\'enez & Lee \& Moser & comm. R. Moser \\
   & (2015) & (2006) & (2015) & \\
Color in figs. & \textcolor{Purple}{$\blacksquare$} & \textcolor{SkyBlue}{$\blacksquare$} & \textcolor{Green}{$\blacksquare$} & \textcolor{SpringGreen}{$\blacksquare$} \\
\label{TableDNS}
\end{tabular}
\end{table}

\subsection{\label{sec32}The outer expansion $U^+_{\mathrm{out}}(Y)$}

\begin{figure}
\center
\includegraphics[width=0.8\textwidth]{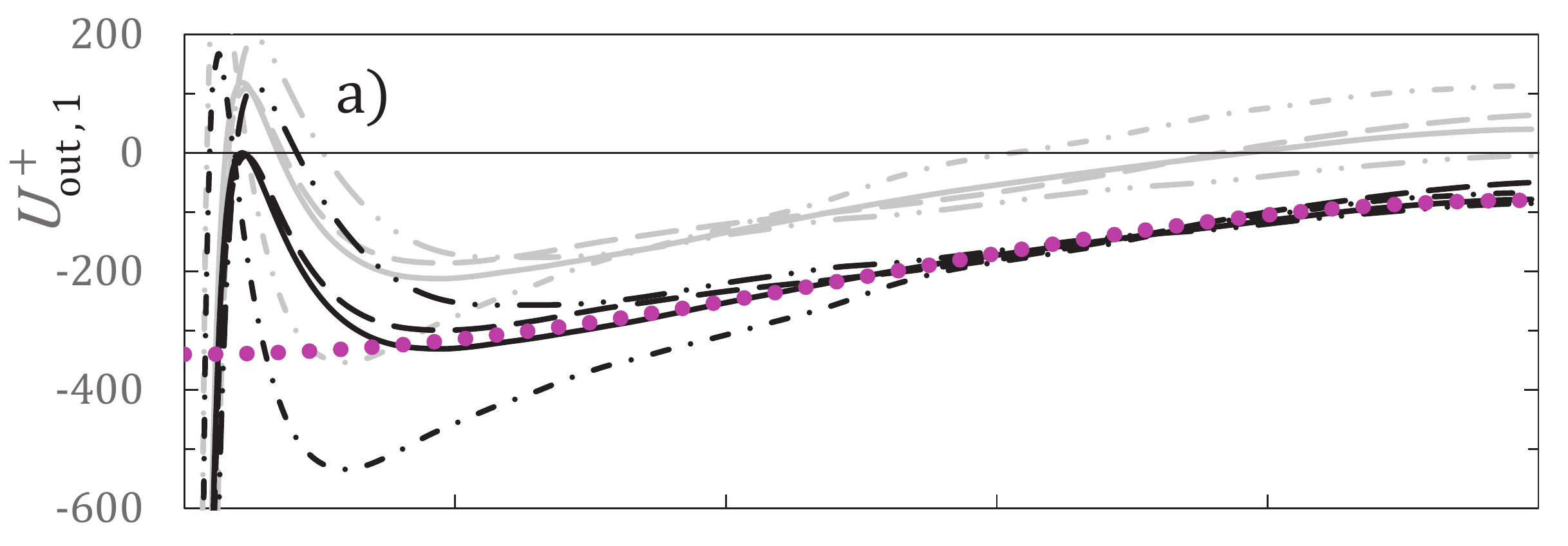}
\includegraphics[width=0.8\textwidth]{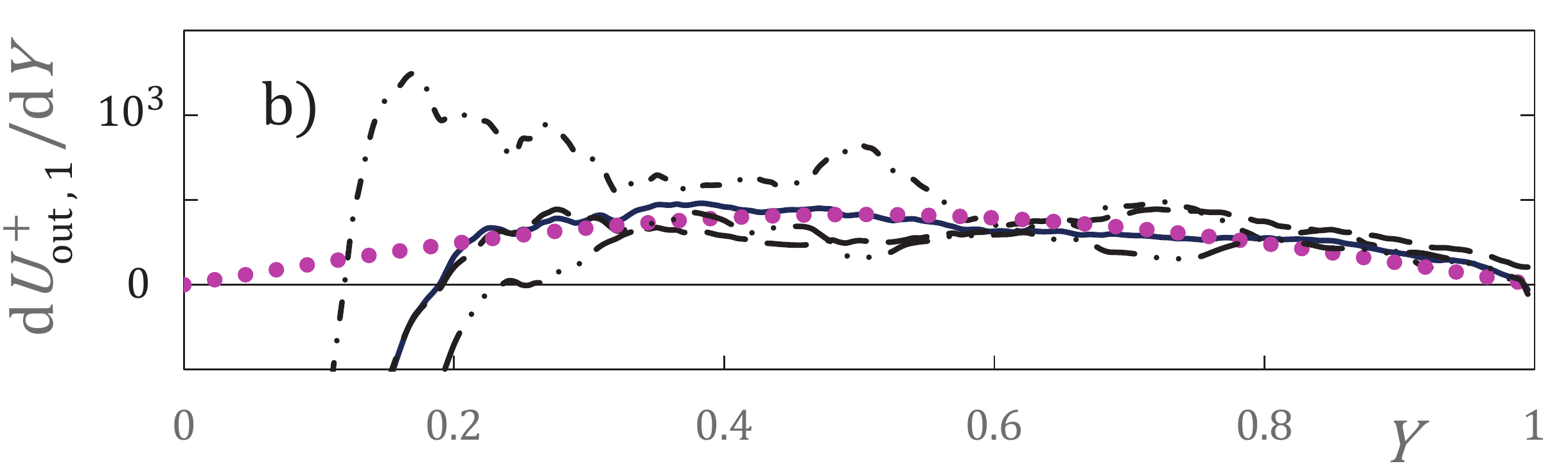}
\caption{(color online)\ (a) Higher order term $U^+_{\mathrm{out}, 1}(Y)$ of the outer expansion for the optimal $\kappa = 0.42$, obtained with \textbf{pairs} of DNS from table \ref{TableDNS} : ---, (\#1,\#3); - - -, (\#1,\#4); $-\cdot -$, (\#1,\#2); $-\cdot\cdot -$, (\#2,\#3); \textcolor{Magenta}{$\bullet \bullet \bullet$}, fit by equ. (\ref{Uout1}); Gray : $U^+_{\mathrm{out}, 1}(Y)$ with same profile pairs, but $\kappa = 0.41$. \quad (b) Derivative $\dd U^+_{\mathrm{out}, 1}(Y)/\dd Y$ obtained from the same DNS pairs as in (a); \textcolor{Magenta}{$\bullet \bullet \bullet$}, derivative of equ. (\ref{Uout1}). }
\label{figUout}
\end{figure}

In the following, $U^+_{\mathrm{out}, 1}(Y)$ is determined with both methods discussed in the introductory part of this section \ref{sec3}. Iterating on $\kappa$, until the best collapse of $U^+_{\mathrm{out}, 1}(Y; \kappa, i, j)$ determined from different profile pairs $(i,j)$ is obtained in the core of the channel, leads to $\kappa = 0.42$.
The resulting good collapse in the region $0.4 \lesssim Y \leq 1$ of the $U^+_{\mathrm{out}, 1}(Y)$, obtained from different profile pairs, is shown in figure \ref{figUout}a, together with the fit
\begin{equation}
\hat{U}^+_{\mathrm{out}, 1} = -210 - 130\,\cos(\pi Y) \sim -340 + \mathcal{O}(Y^2) \quad \mathrm{for} \,\, Y\to 0 \quad ,
\label{Uout1}
\end{equation}
where the reader is reminded that analytical fits are designated by hats, while quantities derived from DNS profiles are left without.

As it turns out, the optimal $\kappa = 0.42$ is rather sharply defined, as seen from the divergence of the $U^+_{\mathrm{out}, 1}(Y)$ for $\kappa=0.41$, obtained from the same profile pairs and included in figure \ref{figUout}a in gray.
The confidence in the fit (\ref{Uout1}) is reinforced by the good match in figure \ref{figUout}b between the derivative of the fit (\ref{Uout1}) and the derivatives $\dd U^+_{\mathrm{out}, 1}/\dd Y$ obtained from the same profile pairs as in figure \ref{figUout}a, with a scheme analogous to equation (\ref{diffin}) that requires no knowledge of $\kappa$.

\begin{figure}
\center
\includegraphics[width=0.8\textwidth]{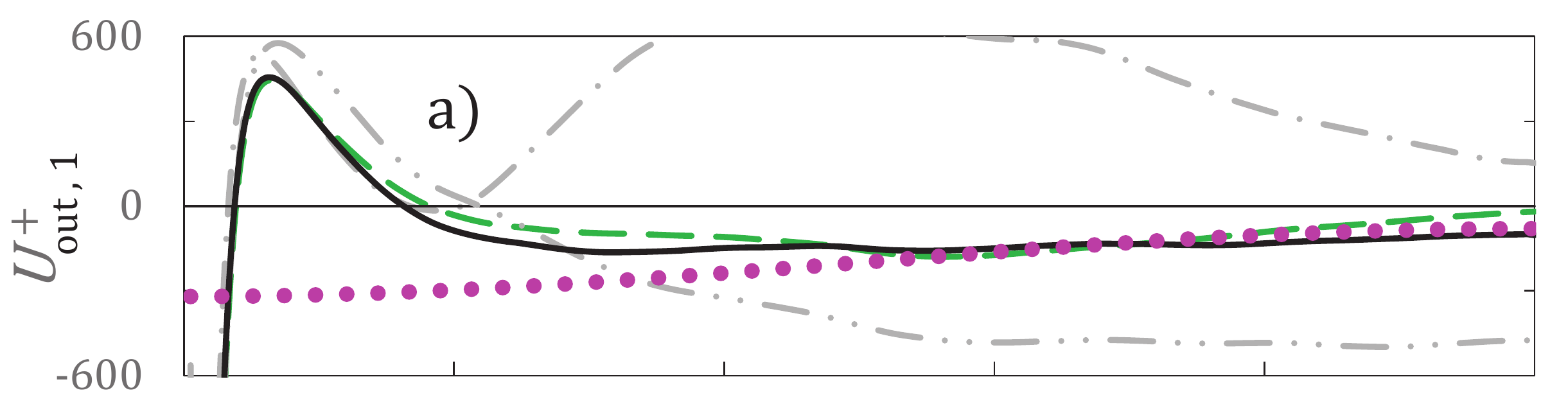}
\includegraphics[width=0.8\textwidth]{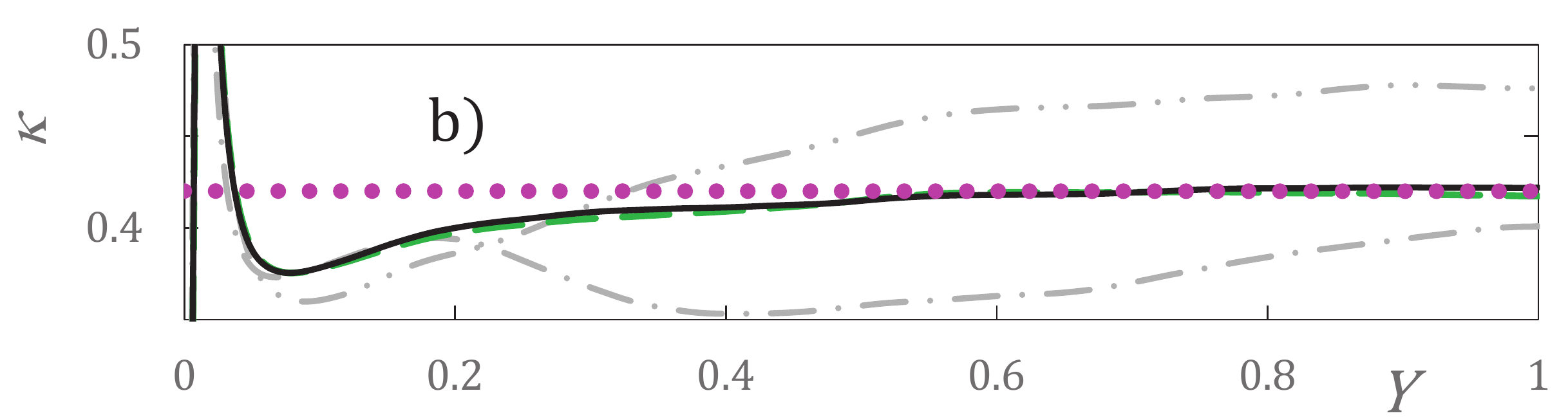}
\caption{(color online)\ (a) Higher order term $U^+_{\mathrm{out}, 1}(Y)$ of the outer expansion obtained from three DNS : ---, (\#1,\#2,\#3); \textcolor{Green}{- - -}, (\#1,\#2,\#4); \textcolor{Gray}{$-\cdot -$}, (\#2,\#3, $\Reytau=3000$ of \citealt{TMG14}); \textcolor{Gray}{$-\cdot\cdot -$}, (\#2,\#3, $\Reytau=4179$ of \citealt{LJ14}); \textcolor{Magenta}{$\bullet \bullet \bullet$}, fit by equ. (\ref{Uout1}). \quad (b) $\kappa$ from the same triplets as in fig. (a); \textcolor{Magenta}{$\bullet \bullet \bullet$}, $\kappa = 0.42$. }
\label{figUout2}
\end{figure}

Since the determination of $U^+_{\mathrm{out}, 1}(Y)$ is a key step of the present analysis, which tests the limits of present DNS, it is useful to compare with the parameter-free method (ii), based on three DNS profiles and outlined in section \ref{sec31}. The resulting $U^+_{\mathrm{out}, 1}(Y)$ is shown in figure \ref{figUout2}a for four profile triplets and the corresponding $\kappa$'s are shown in panel (b). What is striking in this figure, are the surprisingly good results for the two triplets involving only profiles from table \ref{TableDNS}. The results for $\kappa$ in particular, which on the centerline deviate by less than $\pm0.003$ from $0.42$, are outstanding. In contrast, the results from the two triplets not included in table \ref{TableDNS} are useless. They have been included to show why the present methodology fails with a number of DNS profiles : As seen in figure \ref{figUout2}, the two ``bad'' triplets yield results consistent with the two ``good'' ones up to around $Y \approx 0.2$, where they are of no interest for the \textbf{outer} expansion, and become erratic towards the centerline. This suggests an imbalance of computational effort between near-wall and core region, which has been recognized and corrected by the Texas group during the computations for \citet{LM14} (private comm. of Bob Moser and HK Lee).

\begin{figure}
\center
\includegraphics[width=0.8\textwidth]{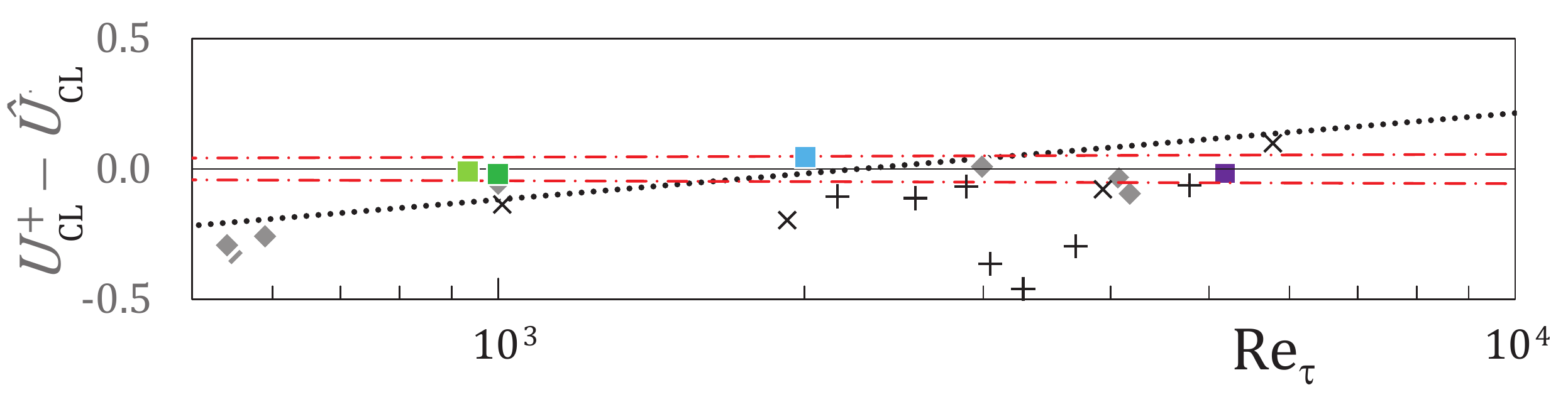}
\caption{(color online)\ Various channel/duct centerline velocities minus $U^+_{\mathrm{CL}}$ (equ. \ref{UCL}) versus $\Reytau$. \textcolor{Purple}{$\blacksquare$} \textcolor{SkyBlue}{$\blacksquare$} \textcolor{Green}{$\blacksquare$} \textcolor{SpringGreen}{$\blacksquare$}, DNS of table \ref{TableDNS}; \textcolor{Gray}{$\blacklozenge$}, other DNS used in \citet{Monk17}; $\times$, \citet{SchultzFlack2013}; $+$, \citet{ZDN03}. \textcolor{red}{$\cdot - \cdot$}, $\pm 0.2\%$ of $U^+_{\mathrm{CL}}$. $\cdot\cdot\cdot$, slope corresponding to $\kappa =0.396$.}
\label{figCL}
\end{figure}

To complete the outer expansion, the leading order term of $U^+_{\mathrm{out}}(Y)$ is split into several contributions
\begin{equation}
U^+_{\mathrm{out}}(Y) = \left\{\frac{1}{0.42}\,\ln\left[\Reytau Y (2-Y)\right] + C + W_0(Y)\right\} + \frac{1}{\Reytau}\,U^+_{\mathrm{out}, 1} + ...\,\, ,
\label{Uouttot}
\end{equation}
where all the terms, including the outer log term, satisfy the channel symmetry $U^+(Y)=U^+(2-Y)$. This requirement is rarely implemented in the literature, where the original decomposition of \cite{Coles56} into simple logarithm and ``wake'' dominates. The only unknowns left in the outer expansion (\ref{Uouttot}) are the constant $C$ and the ``wake function'' $W_0(Y)$ (note that $W_0$ is different from Coles' wake function because of the symmetrized log term). Specifying $W_0(Y=1)=0$ in equation \ref{Uouttot} leads to the 2-term expansion of the centerline velocity
\begin{equation}
\hat{U}^+_{\mathrm{CL}} = \left\{\frac{1}{0.42}\,\ln(\Reytau) + 6.22\right\} - \frac{80}{\Reytau} + \mathcal{O}(\Reytau^{-2})
\label{UCL}
\end{equation}
with the optimal $C=6.22$. Equation (\ref{UCL}) is seen in figure \ref{figCL} to reproduce the reference DNS data of table \ref{TableDNS} with an error of less than $0.2 \%$, which is marginally better than the leading order fit by \citet{Monk17}.

\begin{figure}
\center
\includegraphics[width=0.8\textwidth]{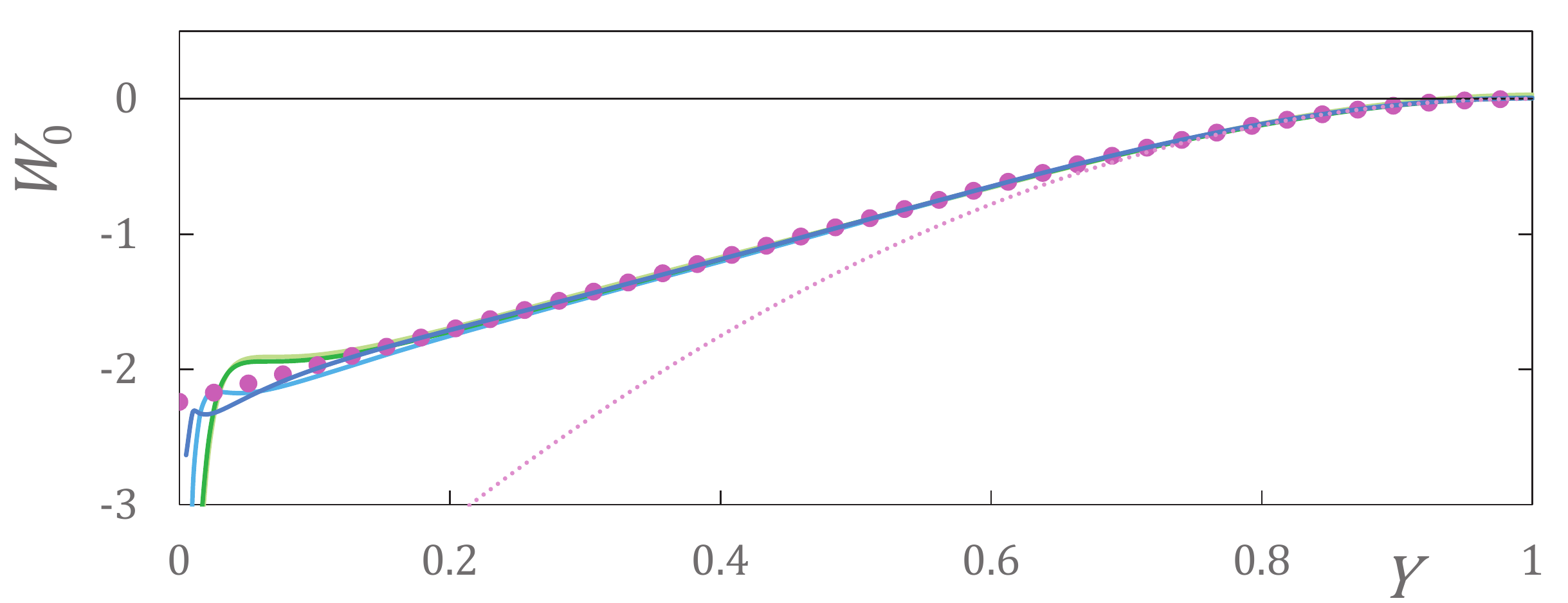}
\caption{(color online)\ $W_0(Y)$ obtained from equ. (\ref{Uouttot}), with $U^+_{\mathrm{out}}(Y)$ approximated by the four DNS profiles of table \ref{TableDNS} (colors as in the table). \textcolor{Magenta}{$\bullet \bullet \bullet$}, fit by equ. (\ref{W0def}); \textcolor{Magenta}{$\cdots$}, leading term $-4.87\,(1-Y)^2$ of Taylor expansion around $Y=1$. }
\label{figUoutW}
\end{figure}

Finally, $W_0$ is obtained from equation (\ref{Uouttot}) by using the fit (\ref{Uout1}) for $U^+_{\mathrm{out}, 1}$ and identifying the total velocity $U^+_{\mathrm{out}}$ with $U^+_{\mathrm{DNS}}$ in the outer region.
The resulting $W_0(Y)$ is shown in figure \ref{figUoutW}, together with its fit
\begin{align}
\frac{\dd\hat{W}_0}{\dd Y} &= 2.66 \, \tanh\left\{3.66\,\frac{1-Y}{[Y(2-Y)]^{2.5}}\right\} \nonumber \\
\hat{W}_0 &= - \int_{Y}^{1} \left[\dd\hat{W}_0/\dd Y\right](Y')\,\dd Y'  \sim  -2.24 + 2.66\,Y + \mathcal{O}(Y^2) \quad \mathrm{for} \,\, Y\to 0 \, ,\label{W0def}
\end{align}
which is necessarily rather elaborate to avoid compromising the determination of the inner expansion in the next section \ref{sec33}.

The two fits $\hat{U}^+_{\mathrm{out}, 1}$ and $\hat{W}_0$ complete the formal description  of the two-term outer expansion (\ref{Uouttot}) of  $U^+$.
For the matching to the inner expansion, to be developed in the next section \ref{sec33}, the limiting behavior of $\hat{U}^+_{\mathrm{out}}$ for $Y\to 0$ is required. After expanding the logarithm in (\ref{Uouttot}) and using the fits (\ref{Uout1}) and (\ref{W0def}), one obtains for $Y \ll 1$
\begin{equation}
\hat{U}^+_{\mathrm{out}}(Y \ll 1) \sim \frac{1}{0.42}\,\ln(\Reytau Y) + 5.63 + 1.47\,Y - \frac{340}{\Reytau} + \mathcal{O}(\Reytau^{-2})\quad ,
\label{Uoutexp}
\end{equation}
where the log-law constant $B=5.63$ is the result of $C=6.22$ minus $2.24$ (equ. \ref{W0def}) plus $\ln(2)/0.42$ from the Taylor expansion of the logarithm in equation (\ref{Uouttot}).

The terms of equation (\ref{Uoutexp}) will appear in the common part, \textbf{only if they have a counterpart in the limit} $y^+\gg 1$ \textbf{of the inner expansion}. As will be seen in the next section \ref{sec33}, this is the case for all the terms in equation (\ref{Uoutexp}).

\begin{figure}
\center
\includegraphics[width=0.49\textwidth]{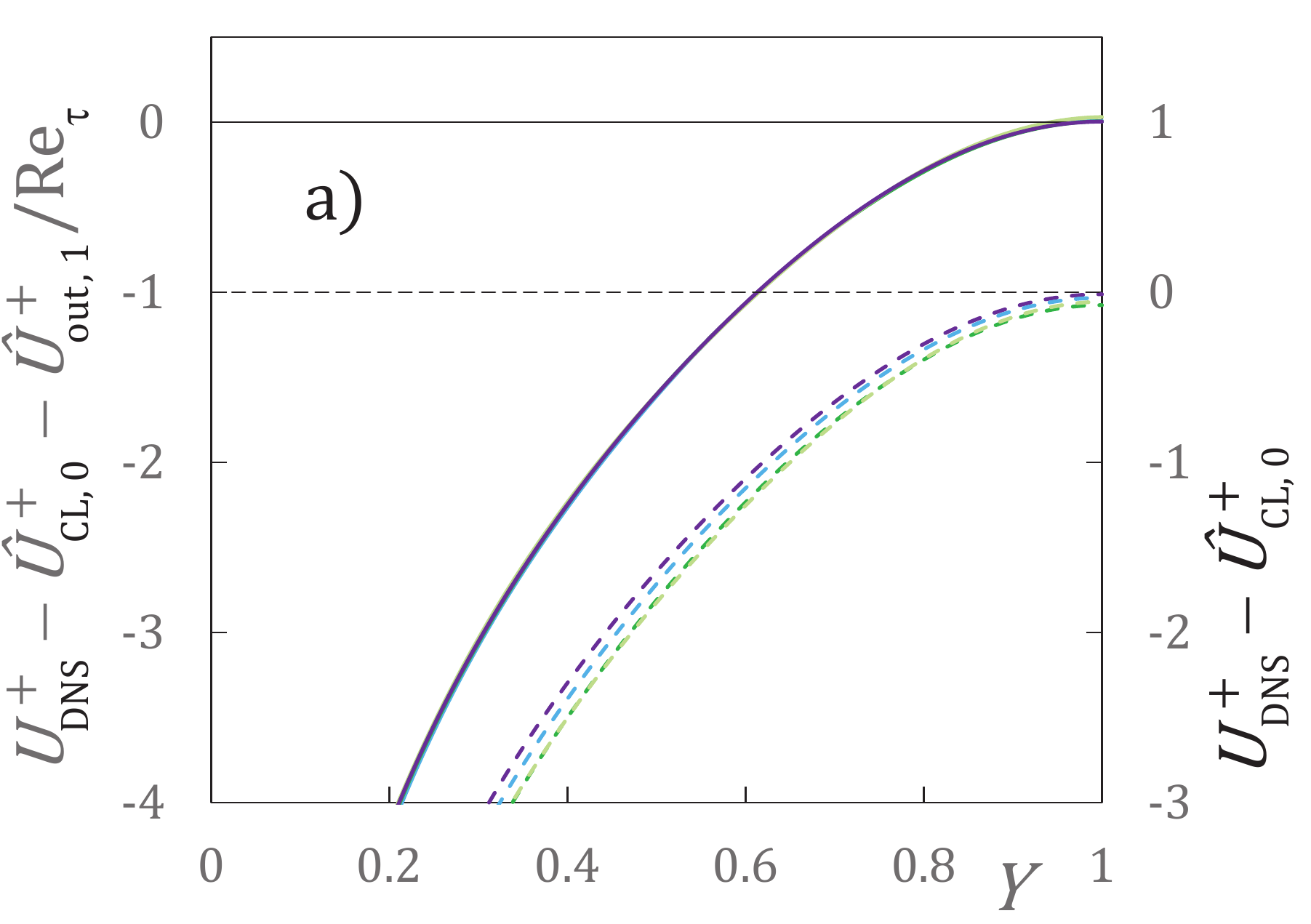}
\includegraphics[width=0.49\textwidth]{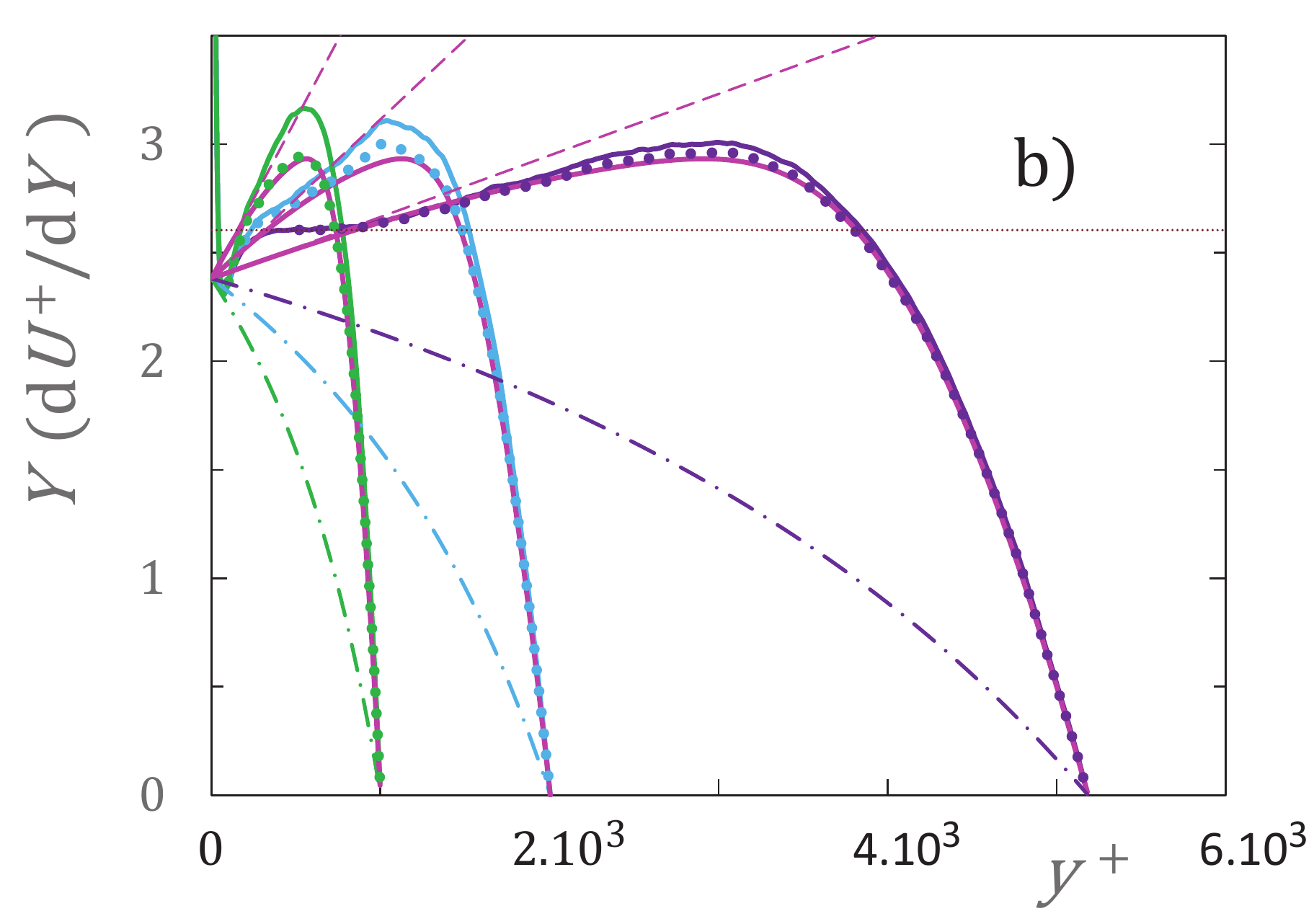}
\caption{(color online)\ (a) The effect of subtracting the fit $\hat{U}^+_{\mathrm{out}, 1}$ (equ. \ref{Uout1}) from the four DNS profiles of table \ref{TableDNS}. $\quad$ (b) solid lines, $Y\,\dd U^+_{\mathrm{DNS}}/\dd Y$ versus $y^+$ for the three highest $\Reytau$ ;  $\bullet \bullet \bullet$, solid lines minus $Y$ times derivative of $\hat{U}^+_{\mathrm{out}, 1}$ (equ. \ref{Uout1}) ; \textcolor{Magenta}{---} , $Y$ times derivative of
leading order outer velocity (equ. \ref{Uouttot}); \textcolor{Magenta}{- - -}, small-$Y$ contribution $(1/0.42)+1.47\,Y$ to $Y\,\dd U^+/\dd Y$ from equ. (\ref{Uoutexp}); \textcolor{Magenta}{$\cdots$}, the apparent plateau (1/0.384) in fig. 3a of \citet{LM14}; $-\cdot - \cdot -$, $Y$ times derivative of logarithm in equ.(\ref{Uouttot}). }
\label{fig4New}
\end{figure}

Before moving on to the inner expansion, it is worthwhile to document in figure \ref{fig4New} the very significant improvement in the description of the outer velocity profile, brought about by the $\mathcal{O}(\Reytau^{-1})$ correction (\ref{Uout1}). Panel (a) documents the improved collapse of the four profiles of table \ref{TableDNS} in the central part of the channel. Panel (b) corresponds to figure 7a of \citet{Jimenez07} and is a striking demonstration of the importance of the linear contribution to the overlap profile (equs. \ref{Uouttot} and \ref{Uoutexp}) and of the $\Reytau^{-1}$ contributions for the interpretation of DNS data at moderate Reynolds numbers and in particular for the determination of $\kappa$. Note however, that the quantity $Y\,\dd U^+_{\mathrm{DNS}}/\dd Y$, chosen as in figure 7a of \citeauthor{Jimenez07} does not have the proper symmetry about the centerline.

\subsection{\label{sec33}The inner expansion $U^+_{\mathrm{in}}(y^+)$ and the final matching}

\begin{figure}
\center
\includegraphics[width=0.8\textwidth]{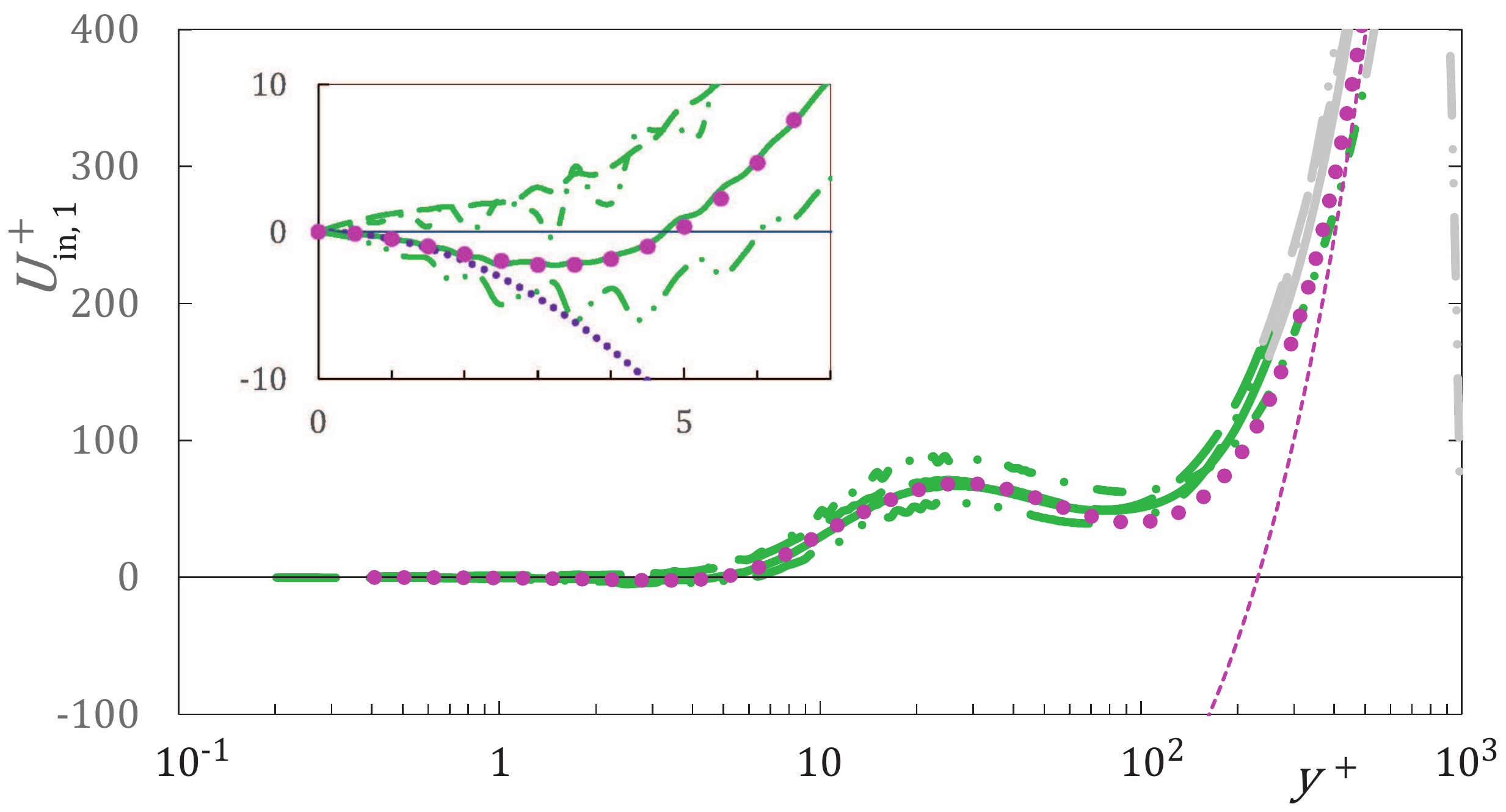}
\caption{(color online)\ First order term $U^+_{\mathrm{in}, 1}(y^+)$ , obtained from differences (\ref{diffin}) of $U^+$-profiles in table \ref{TableDNS}. Profile pairs and line styles as in figure \ref{figUout}a (Green lines up to $Y=0.25$ for the lower $\Reytau$ of the pair, gray lines beyond).  \textcolor{Magenta}{$\bullet \bullet \bullet$} , complete fit $\hat{U}^+_{\mathrm{in,1}}$ by equ. \ref{Uin1}; \textcolor{Magenta}{- - -}, linear function $1.47\,y^+ - 340$ matching the linear part of $\hat{U}^+_{\mathrm{out}}(Y \ll 1)$ (equ. \ref{Uoutexp}). Insert : blowup of the origin with \textcolor{Violet}{$\cdots$} , $- (1/2) (y^+)^2$.}
\label{fig6New}
\end{figure}

Starting again with the order $\mathcal{O}(\Reytau^{-1})$, the first order of the inner expansion $U^+_{\mathrm{in}, 1}(y^+)$ is determined with equation (\ref{diffin}), which is completely parameter-free. Although at the limit of DNS uncertainty, different pairs of the profiles in table \ref{TableDNS} yield reasonably consistent $U^+_{\mathrm{in,1}}$, seen in figure \ref{fig6New} to have three distinct features :
\begin{enumerate}
\item An initial negative excursion near the origin due to the pressure gradient, which produces the exact quadratic term $-(\beta/2)(y^+)^2$ in the Taylor expansion of $U^+$ about the wall. This minute negative part is correctly reproduced by two of the four DNS pairs.
\item A first order ``hump'' very similar to the hump proposed by \citet{NagibChauhan2008} to improve the Musker profile [see appendix \ref{sec:App1} and equation (\ref{Hump})], except that its height diminishes as $\Reytau^{-1}$.
\item A final approach to the linear function $1.47\,y^+ - 340$ which matches the linear part of $\hat{U}^+_{\mathrm{out}}(Y \ll 1)$ (equ. \ref{Uoutexp})\newline
\end{enumerate}

These three distinct features of the 1st order inner velocity are fitted by the three terms of
\begin{align}
\hat{U}^+_{\mathrm{in,1}} &= -\,\frac{1}{2} (y^+)^2\,\exp\left[-0.004\,(y^+)^3\right] + \hat{H}_{\mathrm{NC}}(y^+; 67, 0.75, 27) + \nonumber \\ &+ 490.5\,\ln\cosh\left[2.996\,10^{-3}\,y^+\right] \sim 1.47\,y^+ - 340 \quad \mathrm{for} \,\, y^+\to \infty \,
\label{Uin1}
\end{align}
with $\hat{H}_{\mathrm{NC}}$ the ``hump'' function of equation (\ref{Hump}).
As required, the large $y^+$ limit of $\hat{U}^+_{\mathrm{in,1}}/\Reytau$ matches the corresponding terms in the small $Y$ expansion (\ref{Uoutexp}) of $\hat{U}^+_{\mathrm{out}}$. Hence, the common part of the 2-term inner and outer expansions is
\begin{equation}
\hat{U}^+_{\mathrm{cp}}(Y) = \left\{\frac{1}{0.42}\,\ln(\Reytau Y) + 5.63 + 1.47\,Y\right\} - \frac{340}{\Reytau} + \mathcal{O}(\Reytau^{-2}) \,,
\label{UcpY}
\end{equation}
or equivalently
\begin{equation}
\hat{U}^+_{\mathrm{cp}}(y^+) = \left\{\frac{1}{0.42}\,\ln(y+) + 5.63\right\} +\frac{1}{\Reytau}\,\left\{1.47\,y^+ - 340\right\} + \mathcal{O}(\Reytau^{-2})
\label{Ucpyp}
\end{equation}

\begin{figure}
\center
\includegraphics[width=0.8\textwidth]{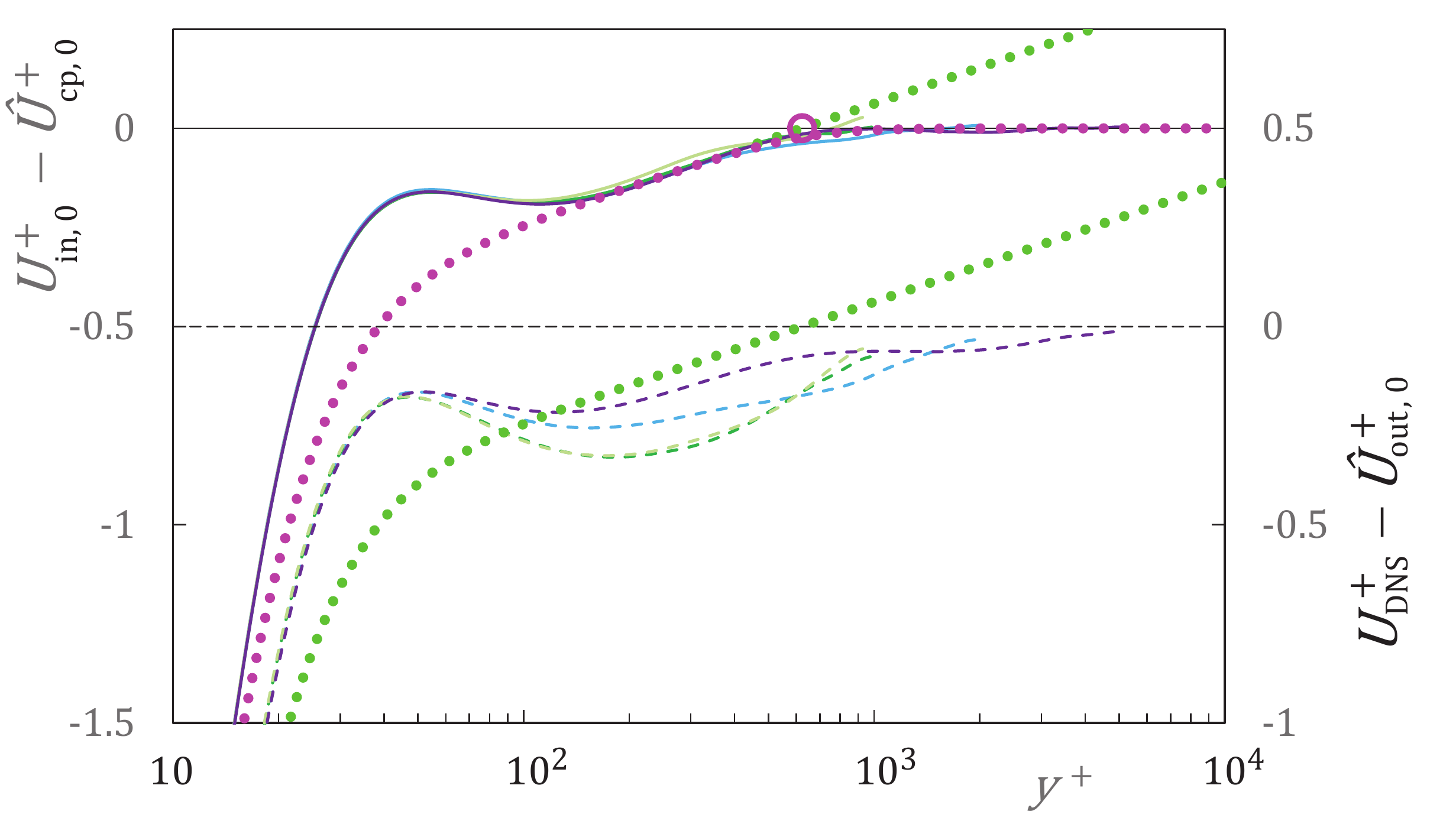}
\caption{(color online) Left axis and solid lines : leading order inner velocity $U^+_{\mathrm{in, 0}}$ minus $U^+_{\mathrm{cp, 0}}(y^+)$ obtained from equ. (\ref{Uin0}) and (\ref{Ucpyp}) for the 4 profiles of table \ref{TableDNS}. Right axis and broken lines : leading order inner velocity minus common part equal to $U^+_{\mathrm{DNS}} - \hat{U}^+_{\mathrm{out,0}}$, determined without the $\mathcal{O}(\Reytau^{-1})$ terms in equ. (\ref{Uin0}). \textcolor{LimeGreen}{$\cdots$}, improved Musker profile $\hat{U}^+_{\mathrm{mM}}$ (equ. \ref{mMusker}) without ``hump'', for $\kappa_{\mathrm{M}}=0.398$ and $B_{\mathrm{M}}=4.784$, minus $\hat{U}^+_{\mathrm{cp},0}(y^+)$  (equ. \ref{Ucpyp});
\textcolor{Magenta}{$\cdots$}, $\hat{U}^+_{\mathrm{mM}}-\hat{U}^+_{\mathrm{cp},0}(y^+) - \hat{\Delta}_{\mathrm{log,Ch}}$, including the change in logarithmic slope (equ. \ref{dellog}), and \textcolor{Magenta}{$\bigcirc$}, the breakpoint at $y^+_{\mathrm{break}}=624$.}
\label{fig7}
\end{figure}

The leading term $U^+_{\mathrm{in,0}}$ of the inner expansion is now finally obtained from the composite expansion
\begin{equation}
U^+_{\mathrm{DNS}} \cong U^+_{\mathrm{in,0}} + \hat{U}^+_{\mathrm{out,0}} - \hat{U}^+_{\mathrm{cp,0}}(y^+)
+ \hat{U}^+_{\mathrm{out,0}} + \frac{1}{\Reytau}\,\left\{\hat{U}^+_{\mathrm{in,1}} + \hat{U}^+_{\mathrm{out,1}} - \hat{U}^+_{\mathrm{cp,1}}(y^+)\right\}\,,
\label{Uin0}
\end{equation}
with the common part expressed in terms of $y^+$, i.e. split into leading and first order parts according to equation (\ref{Ucpyp}). The resulting $U^+_{\mathrm{in,0}}$ minus the leading order common part is shown on the left axis of figure \ref{fig7}. For comparison, the same quantity, obtained without the $\mathcal{O}(\Reytau^{-1})$ terms in equation (\ref{Uin0}), is plotted on the right axis and the striking improvement brought about by taking $\mathcal{O}(\Reytau^{-1})$ terms into account is evident. This improvement also reveals a clean logarithmic region of $U^+_{\mathrm{in,0}}$ beyond $y^+\approx 150$, where the Musker fit has already reached its design log-law with $\kappa_{\mathrm{M}}=0.398$ and $B_{\mathrm{M}}=4.717$. This first logarithmic region ends at a breakpoint $y^+_{\mathrm{break}}=624$ (the magenta circle in fig. \ref{fig7}), where $U^+_{\mathrm{in,0}}$ switches to the true leading-order overlap log-law  $(1/0.42)\,\ln(y^+) + 5.63$ of equation (\ref{Ucpyp}).

\begin{figure}
\center
\includegraphics[width=0.8\textwidth]{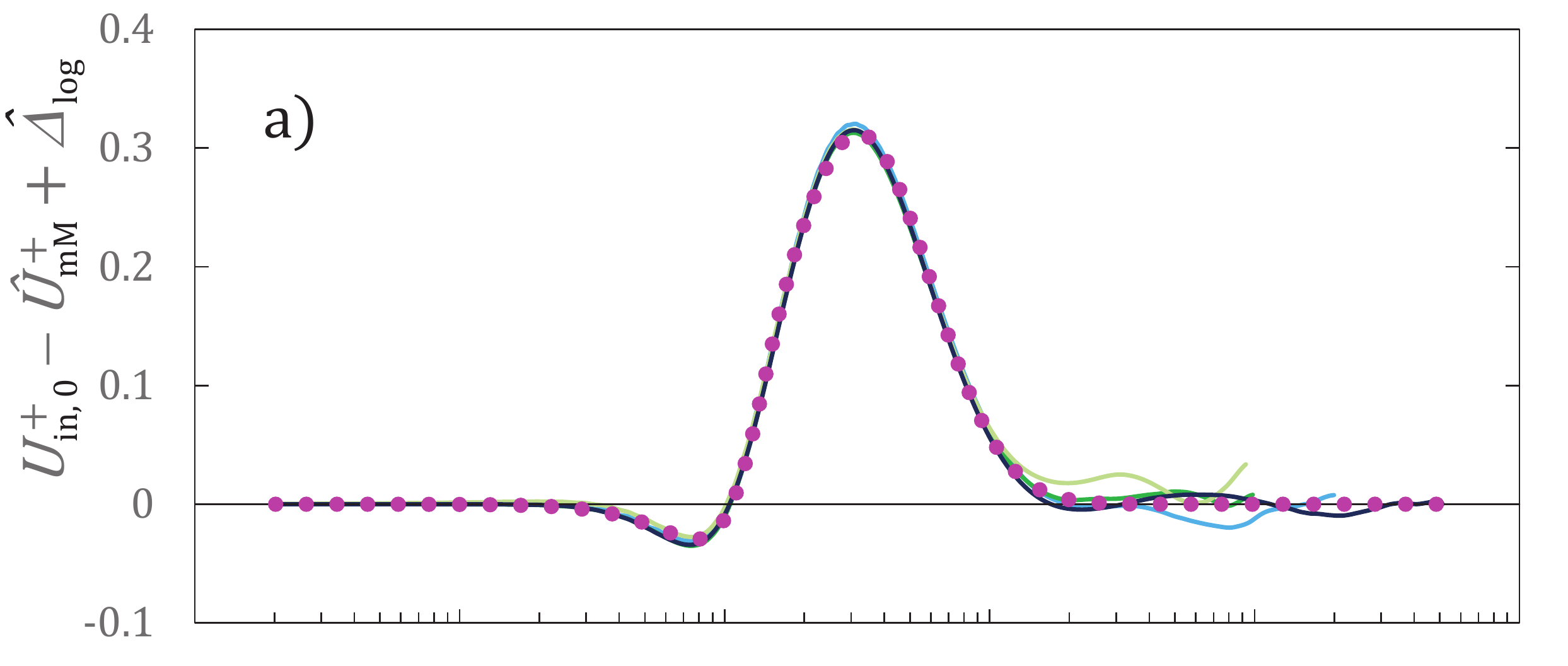}
\includegraphics[width=0.8\textwidth]{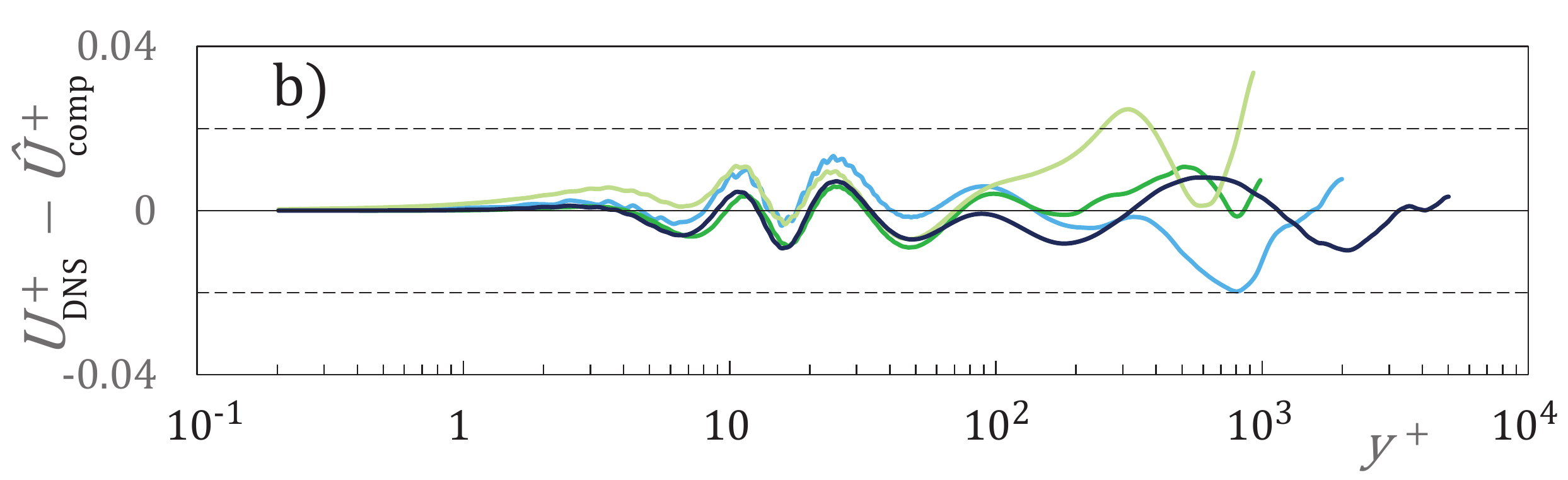}
\caption{(color online) (a) $U^+_{\mathrm{in, 0}} - \hat{U}^+_{\mathrm{mM}} + \hat{\Delta}_{\mathrm{log,Ch}}$ (equ. \ref{mMusker} and \ref{dellog}) for the 4 profiles of table \ref{TableDNS} (same color scheme); \textcolor{Magenta}{$\bullet \bullet \bullet$}, fit by equ. (\ref{Uin1}) (b) DNS profiles $U^+_{\mathrm{DNS}}$ minus complete composite fit $\hat{U}^+_{\mathrm{comp}}$ up to and including $\mathcal{O}(\Reytau^{-1})$ terms. }
\label{fig8}
\end{figure}

This change of logarithmic slope at $y^+_{\mathrm{break}}=624$ is well fitted by the function
\begin{equation}
\hat{\Delta}_{\mathrm{log,Ch}}(y^+) = \frac{1}{5}\,\left[\frac{1}{0.398} - \frac{1}{0.42}\right]\,\ln\left[1 + \left(\frac{y^+}{624}\right)^5\right]
\label{dellog}
\end{equation}
already used by \citet{Monk17}.

The last term of the two-term composite expansion of $U^+$ to be fitted is $U^+_{\mathrm{in, 0}}$. This is achieved in two steps: First, the modified Musker profile (equ. \ref{mMusker}) with $\kappa_{\mathrm{M}}=0.398$ and $B_{\mathrm{M}}=4.784$ is subtracted, and the changeover to the true log law $\hat{\Delta}^+_{\mathrm{log}}$ (equation \ref{dellog}) is added. The result is shown in figure \ref{fig8}a which reveals the leading order hump, seen to be similar to the one discussed by \citet{NagibChauhan2008}. To maintain the highest possible fidelity of the fits, the hump of figure \ref{fig8}a is described by the modified Hump function
\begin{equation}
\hat{H}_{\mathrm{mNC}}(y^+) =  \hat{H}_{\mathrm{NC}}(y^+; 0.313, 1.35, 33) - 2.10^{-4}(y^+)^3 \exp\left[-(0.1 y^+)^3\right] \,,
\label{modhump}
\end{equation}
with the function $\hat{H}_{\mathrm{NC}}$ given by equation (\ref{Hump}).

Putting equations (\ref{mMusker}), (\ref{dellog}) and (\ref{modhump}) together, the complete fit of $U^+_{\mathrm{in, 0}}$ is obtained as
\begin{equation}
\hat{U}^+_{\mathrm{in, 0}} = \hat{U}^+_{\mathrm{mM}}(y^+; 0.398, 4.784) - \hat{\Delta}_{\mathrm{log,Ch}} + \hat{H}_{\mathrm{mNC}}
\label{fitUin0}
\end{equation}
At this point, all the terms of the composite expansion have been fitted, and the complete composite fit can now be compared to the four DNS profiles of table \ref{TableDNS}. The result is shown in figure \ref{fig8}b which demonstrates an unprecedented collapse of all the four DNS profiles onto the composite profile $\hat{U}^+_{\mathrm{comp}}$, with absolute deviations of less than $\pm 0.02$.

In conclusion, this analysis including for the first time all the $\mathcal{O}(\Reytau)^{-1}$ terms has demonstrated that the difference between the Musker $\kappa_{\mathrm{M}}$ and the overlap $\kappa$ is not a statistical accident (see in particular figure \ref{fig7}). Nevertheless, the present findings are based on the restricted set of DNS in table \ref{TableDNS}. Profile pairs from other sources have been considered and have yielded somewhat erratic first order terms especially for the outer expansion, different from the term obtained in section \ref{sec32}. As already mentioned, the likely reason for the differences between DNS is the amount of computational resources devoted to the outer flow.

\section{\label{sec4} Evidence for the hypothesis (\ref{hyp}) in Couette flow and review of pipe DNS}

\subsection{\label{sec41} Couette flow}

\begin{figure}
\center
\includegraphics[width=0.8\textwidth]{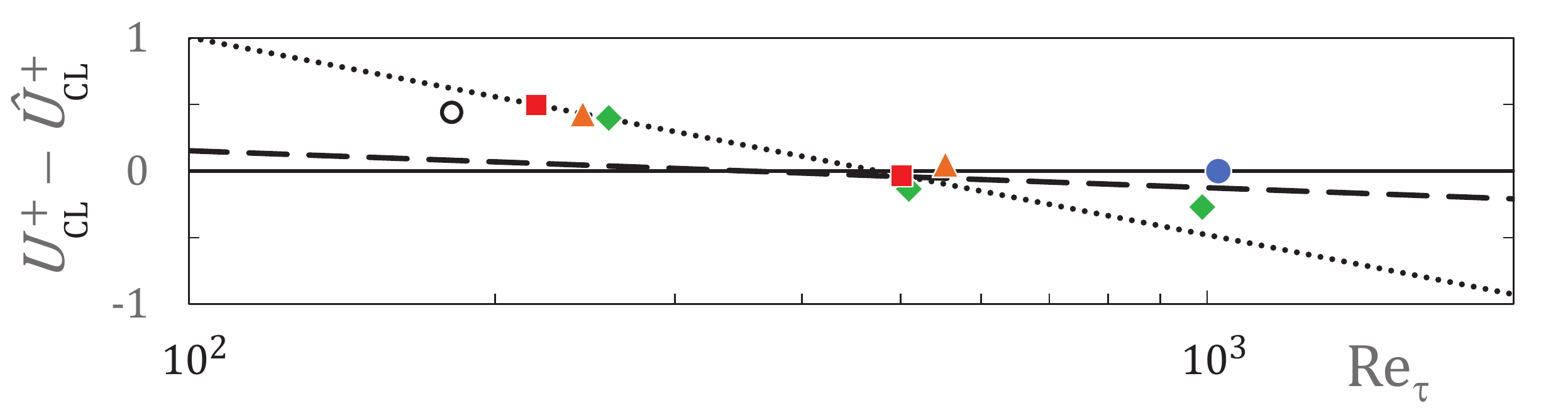}
\caption{(color online)\ Various Couette centerline velocities minus $\hat{U}^+_{\mathrm{CL}}$ (equ. \ref{CUCL}) versus $\Reytau$ from different DNS. $\circ$, \citet{Tsukahara2006}; \textcolor{Red}{$\blacksquare$}, \citet{lee_moser_2018}; \textcolor{Orange}{$\blacktriangle$}, \citet{avsarkisov_etal_2014}; \textcolor{Green}{$\blacklozenge$}, \citet{pirozzoli2014}; \textcolor{Blue}{$\bullet$}, \citet{krah2018}. - - -, $(1/0.384)\ln(\Reytau)+3.75$ minus equ. (\ref{CUCL}); $\cdot\cdot\cdot$, $(1/0.481)\ln(\Reytau)+7.01$ minus equ. (\ref{CUCL}).}
\label{figCouCL}
\end{figure}

The main test of the hypothesis (\ref{hyp}) consists of the demonstration, that in Couette flow the logarithmic slope of $U^+$ increases at $y^+_{\mathrm{break}}$. In terms of an ``eddy'' model, this increase of logarithmic slope in the region $y^+\geq y^+_{\mathrm{break}}$ is brought about by eddies, which originate from the opposite wall, weaken progressively and stop contributing to the mean shear at $y^+_{\mathrm{break}}$. In view of the low $\Reytau$'s of the available DNS, it would be desirable to construct inner and outer asymptotic expansions to order $\mathcal{O}(\Reytau)^{-1}$ analogous to those for the channel in section \ref{sec3}. However, as seen in figure \ref{figCouCL}, the available DNS are limited to $\Reytau\lessapprox 10^3$ and there are doubts on whether the turbulence is fully developed below $\Reytau$ of 500. Furthermore, several of the authors cited in the caption of figure \ref{figCouCL} report, that even the mean velocity profile is sensitive to both the stream-wise and span-wise size of the computational box.

Therefore, only the most recent profile of \citet{krah2018} for the highest $\Reytau=1026$, obtained with state of the art numerical methods, is analyzed here. This, of course, limits the analysis to the leading order of inner and outer expansions and leaves some uncertainty about the expansion parameters.

To start the analysis, the centerline $\kappa$ and $C$ are determined from the data points of \citet{lee_moser_2018} at $\Reytau=501$ and \citet{krah2018} at $\Reytau=1026$ in figure \ref{figCouCL}:
\begin{equation}
\hat{U}^+_{\mathrm{CL}} = \frac{1}{0.367}\,\ln(\Reytau) + 3.04 \quad ,
\label{CUCL}
\end{equation}

Analogous to equation (\ref{Uouttot}) for the channel, except for the opposite symmetry of the logarithm about the centerline, the leading order outer velocity is described by
\begin{equation}
\hat{U}^+_{\mathrm{out}}(Y) = \frac{1}{0.367}\,\ln\left[\frac{\Reytau\,Y}{2-Y}\right] + 3.04 + W_0(Y) \,\, \mathrm{with} \,\,
W_0(Y)= 2.15\,\cos\left(\frac{\pi}{2}\,Y\right)
\label{CUout}
\end{equation}
\begin{equation}
\hat{U}^+_{\mathrm{out}}(Y\ll 1) \sim \frac{1}{0.367}\,\ln(\Reytau\,Y) + 3.30 + 1.36\,Y + \mathcal{O}(Y)^2
\label{CUoutY}
\end{equation}
Since the linear term in the small-$Y$ limit (\ref{CUoutY}) translates in the inner expansion to a first order term, not considered here, the common part consists of just the log-law in equation (\ref{CUoutY}):
\begin{equation}
\hat{U}^+_{\mathrm{cp}} = \frac{1}{0.367}\,\ln(y^+) + 3.30
\label{CUCP}
\end{equation}

\begin{figure}
\center
\includegraphics[width=0.8\textwidth]{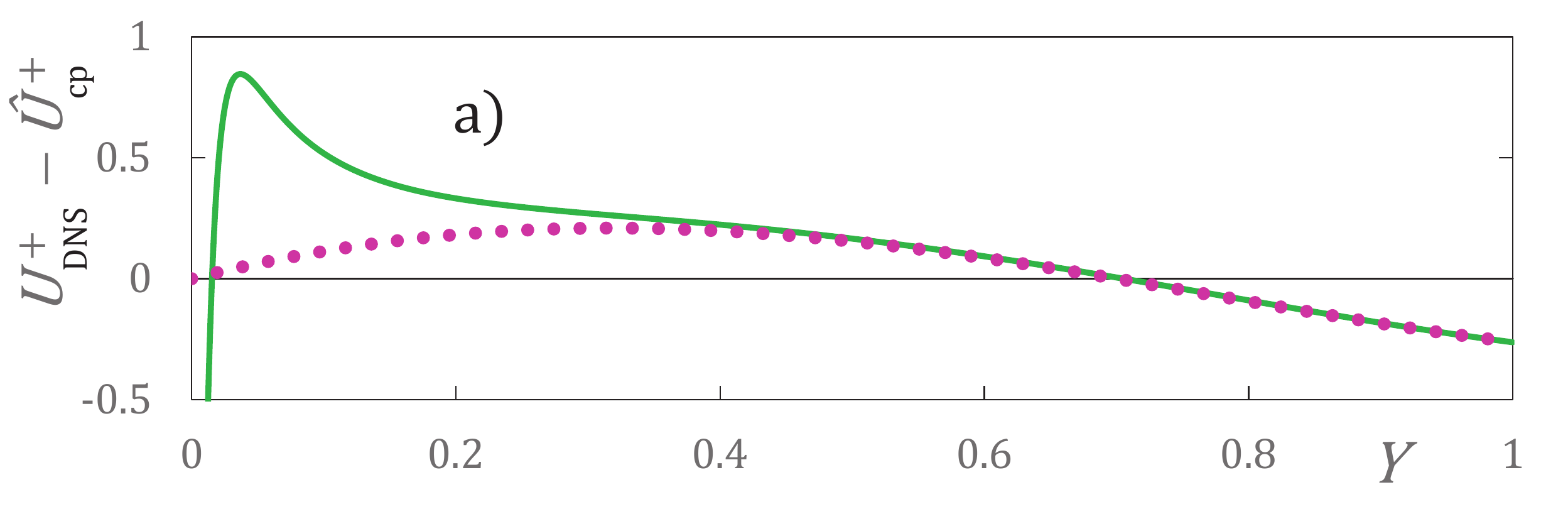}
\includegraphics[width=0.8\textwidth]{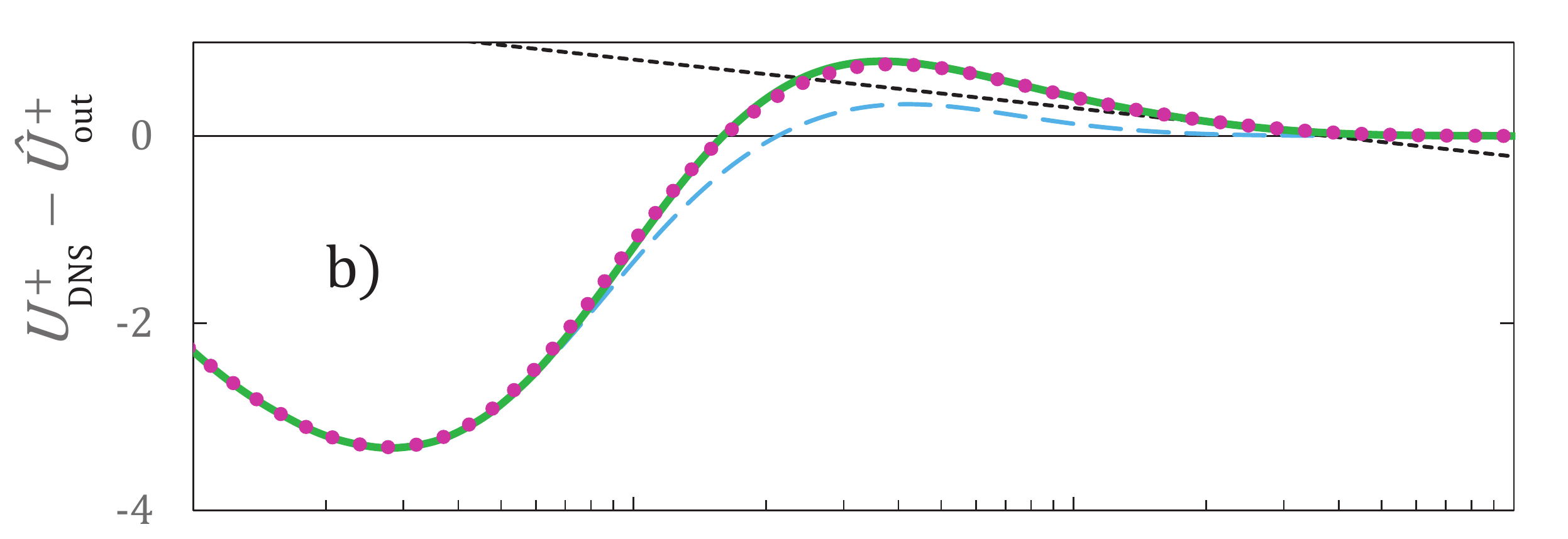}
\includegraphics[width=0.8\textwidth]{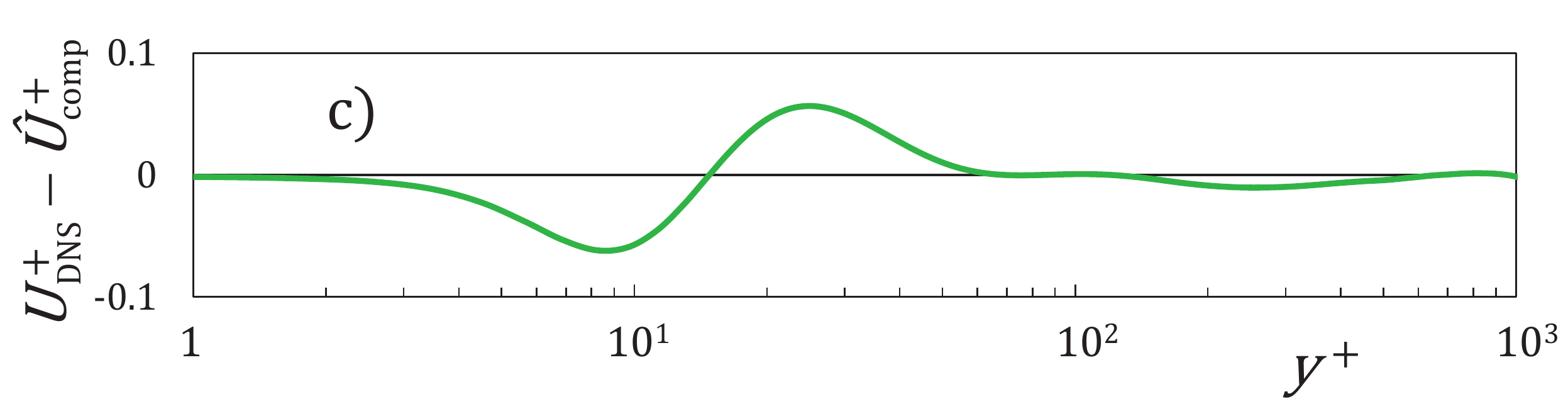}
\caption{(color online)\, (a) \textcolor{Green}{\textbf{---}}, difference between the $U^+$-profile of \citet{krah2018} for $\Reytau =1026$ and $\hat{U}^+_{\mathrm{CP}}$ (equ. \ref{CUCP}); \textcolor{Magenta}{$\bullet\bullet\bullet$}, outer fit $\hat{U}^+_{\mathrm{out}}(Y)$ (equ. \ref{CUout}) minus $\hat{U}^+_{\mathrm{CP}}$ (equ. \ref{CUCP}).
\quad (b) \textcolor{Green}{\textbf{---}}, $U^+_{\mathrm{DNS}}(y^+)$ minus the outer fit $\hat{U}^+_{\mathrm{out}}$ (equ. \ref{CUout}); \textcolor{Aquamarine}{\textbf{- - -}}, $\hat{U}^+_{\mathrm{mM}}(y^+; 0.367, 3.30) + \hat{H}_{\mathrm{NC}}(y^+; 0.38, 1, 34)$ (equs. \ref{mMusker} and \ref{Hump}); \textcolor{Magenta}{$\bullet\bullet\bullet$}, $\hat{U}^+_{\mathrm{in}} - \hat{U}^+_{\mathrm{cp}}$ (equs. \ref{fitUin02} and \ref{CUCP}); - - -, asymptote of $\hat{U}^+_{\mathrm{mM}}(y^+; 0.40, 4.64)$.
\newline (c) \textcolor{Green}{\textbf{---}}, $U^+_{\mathrm{DNS}}(y^+)$ minus the composite fit $\hat{U}^+_{\mathrm{in}} + \hat{U}^+_{\mathrm{out}} - \hat{U}^+_{\mathrm{cp}}$.}
\label{figCou}
\end{figure}

The simple wake function $W_0(Y)$ is seen in panel (a) of figure \ref{figCou} to provide an excellent fit to the DNS for $Y \gtrapprox 0.4$, which is all that can be expected at this low Reynolds number. Panel (b) of figure \ref{figCou} shows the same $U^+_{\mathrm{DNS}}(y^+)$ minus the outer fit $\hat{U}^+_{\mathrm{out}}$ of equation (\ref{CUout}), which is equal to $\hat{U}^+_{\mathrm{in}}-\hat{U}^+_{\mathrm{cp}}$, up to terms of order $\mathcal{O}(\Reytau)^{-1}$. Fitting $\hat{U}^+_{\mathrm{in}}$ with the modified Musker profile plus hump, $\hat{U}^+_{\mathrm{mM}}(y^+; 0.367, 3.30) + \hat{H}_{\mathrm{NC}}(y^+; 0.38, 1, 34)$ (equs. \ref{mMusker} and \ref{Hump} with the overlap parameters of equ. \ref{CUCP}), is seen in figure \ref{figCou}b to be obviously inadequate.
A proper fit of $U^+_{\mathrm{in}}$ is only possible with $\kappa_\mathrm{M}=0.40$, requiring a change of logarithmic slope at  $y^+_{\mathrm{break}}=379$, described by the function
\begin{equation}
\hat{\Delta}_{\mathrm{log,Cou}}(y^+) = \frac{1}{4}\,\left[\frac{1}{0.40} - \frac{1}{0.367}\right]\,\ln\left[1 + \left(\frac{y^+}{379}\right)^4\right] \quad ,
\label{dellog2}
\end{equation}
similar to $\hat{\Delta}_{\mathrm{log,Ch}}$ of equation (\ref{dellog}).

Hence, the leading order of the inner expansion for the mean velocity in Couette flow is
\begin{equation}
\hat{U}^+_{\mathrm{in}} = \hat{U}^+_{\mathrm{mM}}(y^+; 0.40, 4.64) + \hat{H}_{\mathrm{NC}}(y^+; 0.38, 1, 34) - \hat{\Delta}_{\mathrm{log,Cou}}(y^+)
\label{fitUin02}
\end{equation}
with the different terms given by equations (\ref{mMusker}), (\ref{Hump}) and (\ref{dellog2}), respectively. The final composite profile is then simply obtained by combining equations (\ref{CUout}), (\ref{CUCP}) and (\ref{fitUin02}). The result is seen in figure \ref{figCou}c to provide an excellent description of the DNS profile of \citet{krah2018}, except for the ``wiggle'' between $y^+ \approx 2$ and 70, which is due to the simple hump model of equation (\ref{Hump}). No attempt has been made to improve the hump fit as for the channel, because the central point - the demonstration of a switch from a lower (1/0.40) to a higher (1/0.367) logarithmic slope at $y^+_{\mathrm{break}}\approxeq 400$ - is not affected.

\subsection{\label{sec42} The prospects of extracting asymptotic expansions from pipe DNS}

Of the three flows considered in this paper, pipe flow is by far the one of largest practical interest and it would therefore be highly desirable to develop a complete asymptotic description of its mean velocity profile from DNS, analogous to the one for the channel in section \ref{sec3}.

However, the three DNS data points, included as red squares in figure \ref{Fig:PCL} do not even provide a consistent fit for the centerline velocity. Looking at the graphs of $\pm(10^3/\Reytau)$, added to the figure to guide the eye, it appears that the coefficient of $\Reytau^{-1}$ on the centerline must be less than $10^3$. However, it is not clear from the present data, whether the higher order correction is positive, negative or zero - clearly, more high accuracy DNS profiles in the $\Reytau$ range of $10^3 - 5.10^3$ are needed to clarify the situation.

\begin{figure}
\center
\includegraphics[width=0.7\textwidth]{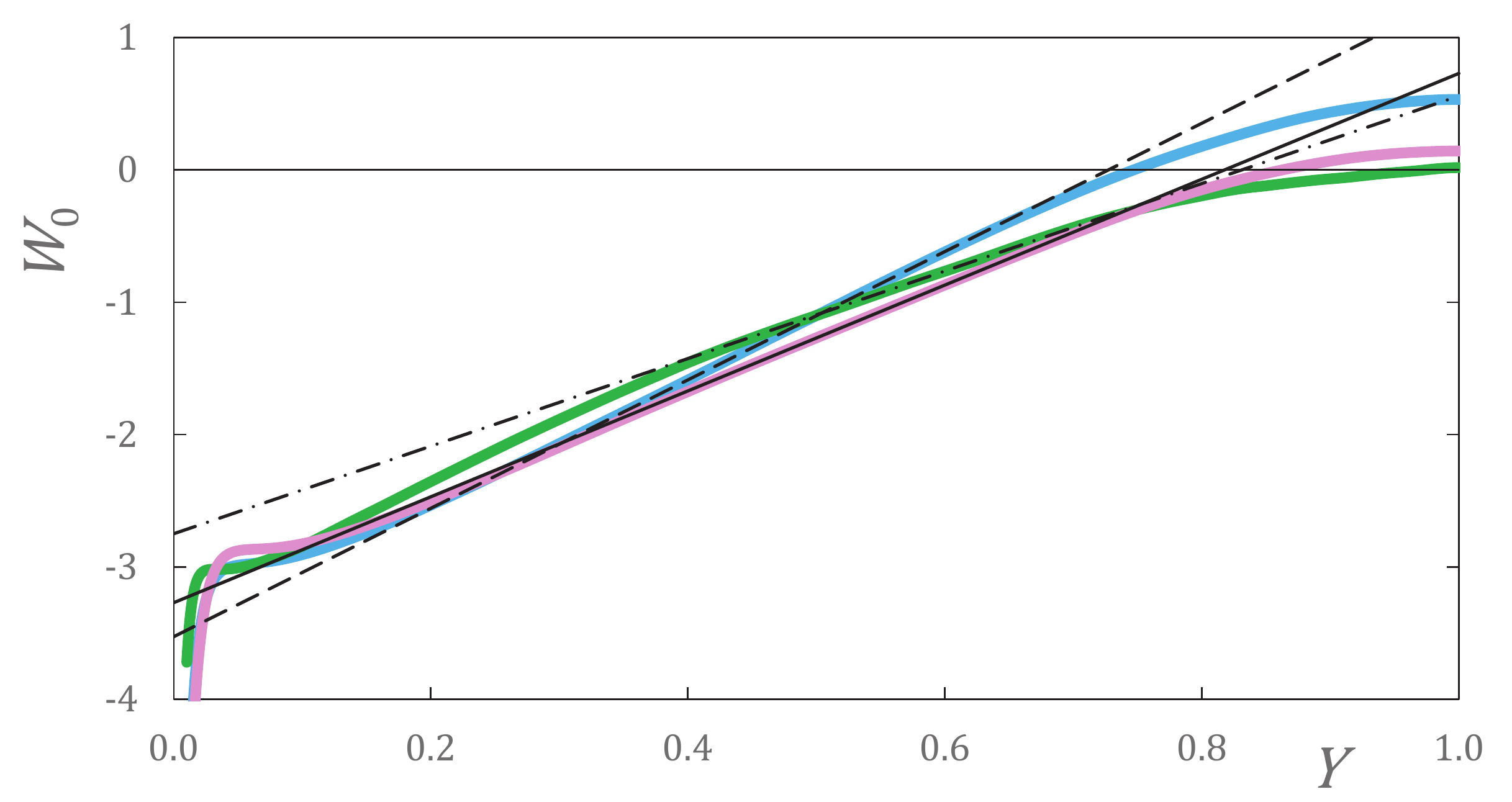}
\caption{\label{Fig:PipeW0} Pipe analogue to figure \ref{figUoutW} with $W_0$ of equ. (\ref{W0pipe}), for the three pipe DNS of figure \ref{Fig:PCL}: \textcolor{Lavender}{---}, $\Reytau=999$, \textcolor{Aquamarine}{---}, $\Reytau=1142$ and \textcolor{Green}{---}, $\Reytau=2003$. ---, - - -, $- \cdot -$, corresponding tentative linear fits with slopes 4.0, 4.9 and 3.3, respectively.}
\end{figure}

To show the necessary improvements in order to perform an analysis analogous to the one in section \ref{sec3} for the channel, the DNS profiles minus the full outer logarithmic part
\begin{equation}
W_0(Y) = U^+_{\mathrm{DNS}} - \left\{(0.42)^{-1}\ln[\Reytau Y(2-Y)] + 6.84\right\}
\label{W0pipe}
\end{equation}
are shown in figure \ref{Fig:PipeW0} for the three pipe DNS of figure \ref{Fig:PCL}. Comparing to figure \ref{figUoutW}, it is obvious that the core region and in particular the handling of the coordinate singularity on the centerline require more attention, before an analysis analogous to section \ref{sec3} can be envisioned. In conclusion, the extraction of asymptotic expansions from DNS (or experiment) analogous to the channel is not yet feasible, and one is left with considerable uncertainty about $\kappa_\mathrm{M}$, $y^+_{\mathrm{break}}$ and $\kappa$ for the pipe.

\section{\label{sec5} Conclusions}

The present investigation, in particular the determination of the two-term inner and outer asymptotic expansions for the channel, together with the Superpipe data and a detailed analysis of the leading order asymptotic expansions for a Couette flow DNS, has uncovered a common feature of ducted parallel flows: a change of logarithmic slope of the mean velocity $U^+(y^+)$ at a $y^+_{\mathrm{break}}$ of several hundred plus units. According to the hypothesis (\ref{hyp}), this slope change depends on the flow symmetry, with slope decrease in channel and pipe flows and an increase in Couette flow.

The evidence for this phenomenon is strong for channel flow, due to a consistent set of DNS, analyzed in section \ref{sec3}.
For Couette flow, the analysis of the DNS profile at the highest available $\Reytau$ supports the hypothesis (\ref{hyp}), but more DNS for $\Reytau$ in the range $10^3 - 5.10^3$ are needed to confirm the present findings and reduce the parameter uncertainty.
For pipe flow, finally, the DNS data base does not yet allow an analogous asymptotic analysis to narrow down the estimates of $\kappa_\mathrm{M}$, $\kappa$ and $y^+_{\mathrm{break}}$, obtained from the Superpipe data.\newline

\noindent The paper concludes with the following list of observations and open questions :
\begin{enumerate}
\item The hight of the ``hump'' above the Musker profile (equ. \ref{Hump}) around $y^+$ of 30 is clearly dependent on $\Reytau$, as shown in section \ref{sec3} for channel flow.  The maximum overshoot over the Musker log-law $\kappa_\mathrm{M}^{-1}\ln(y^+)+B_\mathrm{M}$ is approximately $0.19 + (65/\Reytau)$, which resolves the discrepancies between the height originally proposed by \citet{NagibChauhan2008} and the fits of low Reynolds number data by \citet{Luchini18}.
\item As discussed in section \ref{sec1}, the linear term $\lambda Y$ in the small-$Y$ limit of $U^+_{\mathrm{out,0}}$, the leading order of the outer expansion, appears in the common part only if the inner expansion is carried to $\mathcal{O}(\Reytau)^{-1}$, i.e. contains a matching term $\lambda y^+/\Reytau$. Most recently, \citet{Luchini17} has claimed that $\lambda$ is equal to the pressure gradient parameter $\beta$, which is equal to 0, 1 and 2 for Couette, channel and pipe flow, respectively. From the present profile analyses, $\lambda=1.36$ for Couette flow (equ. \ref{CUoutY}), $\lambda=1.47$ for channel flow (equ. \ref{Uoutexp}) and $\lambda\in [2.1, 3.7]$ for pipe flow, where these latter values correspond to the range of estimated slopes of the three $W_0$ (equ. \ref{W0pipe}) in figure \ref{Fig:PipeW0}, minus $(2\,\kappa)^{-1}$ from the small-$Y$ expansion of $\kappa^{-1}\ln(2-Y)$. The reason for this discrepancy is exposed in appendix \ref{App2}.
\item The analysis of both channel flow in section \ref{sec3} and Couette flow in section \ref{sec41}, found for the inner logarithmic region between $y^+\approxeq 150$ and $y^+_{\mathrm{break}}$ a $\kappa_\mathrm{M}$ of essentially 2/5. However, the question whether this value is possibly universal for truly one-dimensional flows will have to wait for more high quality pipe and Couette DNS, allowing the extraction of precise asymptotic expansions.
\item Once the overlap $\kappa$'s for pipe and Couette flows are finally ``nailed down'' with an uncertainty below say $\pm 0.005$, one should be able to answer the question, whether $\kappa$ is a function of the pressure gradient, or rather depends on the symmetry $S$ of mean vorticity about the centerline, with $S=-1$ for channel and pipe, and $S=1$ for Couette flow. A relation such as $\kappa \approxeq \kappa_\mathrm{M}+0.03\,S$ appears compatible with the profiles analyzed in this paper.
\item The sudden decrease of logarithmic slope at $y^+_{\mathrm{break}}\approxeq 600$, found in the channel for the mean velocity $U^+$ (see section \ref{sec33}), appears also in the profiles of fluctuating pressure $\langle p'p'\rangle^+$ obtained by \citet{Panton2017}. This is particularly evident for the pressure derived from the $\Reytau=5186$ data of \citet{LM14}  (profile \#1 in table \ref{TableDNS}): in figure 1a of Panton et al. for the total $\langle p'p'\rangle^+$, the decrease of slope from the interval $y^+\in [150, 600]$ to $y^+>600$ is perceptible, but not very marked. However, in their figure 3 the pressure indicator function for the highest $\Reytau=5186$ clearly shows two plateaus with a decrease of logarithmic slope by around 5\% between $y^+$ of 500 and $10^3$. This slope change is also evident in their figure 4a for the ``rapid pressure''. As noted by \citet{Panton2017}, the near-correspondence between the logarithmic slopes in the $\langle p'p'\rangle^+$ profiles for the channel and the ones obtained in section \ref{sec3} from the asymptotic analysis of $U^+$ is unexplained and begs for further research.
\item Finally, it might also be interesting to determine the distribution of eddies attached to, or originating from the opposite wall, which preserves the logarithmic law in the overlap layer beyond $y^+_{\mathrm{break}}$ \citep[for the attached eddy model, see e.g.][and references therein]{MarusicMonty19}.
\newline
\end{enumerate}

\begin{acknowledgments}
I am grateful to my partner for bearing with me while staring at graphs for days on end. Thanks also to Ron Panton for pointing out the connections to the fluctuating pressure profiles and to Marco Giometto for useful comments on the draft of this paper.
\end{acknowledgments}

Declaration of Interests. The author reports no conflict of interest.
\appendix
\section{\label{sec:App1}The ``Musker'' fit for the inner $U^+$-profile, with additions}

The Musker profile \citep{Musker79} is obtained by integrating $\dd U^+/\dd y^+ =[\kappa_\mathrm{M} S + (y^+)^2][\kappa_\mathrm{M} S + (y^+)^2 + \kappa_\mathrm{M} (y^+)^3]^{-1}$ analytically, where the subscript $M$ designates parameters used to generate the Musker fit. Note in particular, that $\kappa_\mathrm{M}$ is not necessarily equal to the overlap $\kappa$. The two parameters $\kappa_\mathrm{M}$ and $S$ determine the asymptotic behavior of the Musker profile $\hat{U}^+_{\mathrm{M}} \sim y^+/\kappa_\mathrm{M} + B_{\mathrm{M}}$ but, as noted by \citet{NagibChauhan2008}, the straightforward integration is prone to numerical near-cancellations. The problem is avoided by recasting the result in the following form :
\begin{align}
\label{Musker}
&\hat{U}^+_{\mathrm{M}}\left(y^+; \kappa_\mathrm{M}, B_\mathrm{M}\right) = \Gamma_1\, \ln\left(1 - \frac{y^+}{g_1}\right) + \frac{\Gamma_2}{2}\,\ln\left(1 - \frac{g_2\, y^+}{g_3} + \frac{(y^+)^2}{g_3}\right) \nonumber \\
&+\frac{2\,\Gamma_3 + \Gamma_2\,g_2}{g_4}\,\left[\arctan\left(\frac{2\,y^+- g_2}{g_4}\right) + \arctan\left(\frac{g_2}{g_4}\right)\right]
\end{align}
with
\begin{align}
&s_{1,2} = \left(-\,\frac{S}{2}\right)^{1/3}\,\left\{1 + \frac{2}{S\,(3\,\kappa_\mathrm{M})^3} \pm \left[1 + \frac{4}{S\,(3\, \kappa_\mathrm{M})^3} \right]^{1/2}\right\}^{1/3} \label{M1}\\
&g_1=s_1+s_2-\frac{1}{3\,\kappa_\mathrm{M}}\,;\quad g_2=-g_1-\frac{1}{\kappa_\mathrm{M}}\,; \nonumber \\
&g_3=\frac{1}{4}\,\left(s_1+s_2+\frac{2}{3\,\kappa_\mathrm{M}}\right)^2 + \frac{3}{4}\,\bigg(s_1-s_2\bigg)^2\,;\quad g_4=\left(4 g_3-g_2^2\right)^{1/2} \label{M2}\\
&\Gamma_1=\frac{S+\kappa_\mathrm{M}^{-1}g_1^2}{g_1^2-g_1g_2+g_3}\,;\quad \Gamma_2=\frac{1}{\kappa_\mathrm{M}}-\Gamma_1\,;\quad \Gamma_3=\frac{g_3\Gamma_1-S}{g_1} \label{M3}
\end{align}
The additive log-law constant $B_\mathrm{M}$ is the limit $y^+\to\infty$ of equation (\ref{Musker})
\begin{equation}
\label{Bloglaw}
B_\mathrm{M} = -\Gamma_1\, \ln\left(-g_1\right) - \frac{\Gamma_2}{2}\,\ln\left(g_2\right) + \frac{2\,\Gamma_3 + \Gamma_2\,g_2}{g_4}\,\left[\frac{\pi}{2} + \arctan\left(\frac{g_2}{g_4}\right)\right]
\end{equation}
and its desired value is obtained by a simple iteration on $S$.

\begin{figure}
\center
\includegraphics[width=0.8\textwidth]{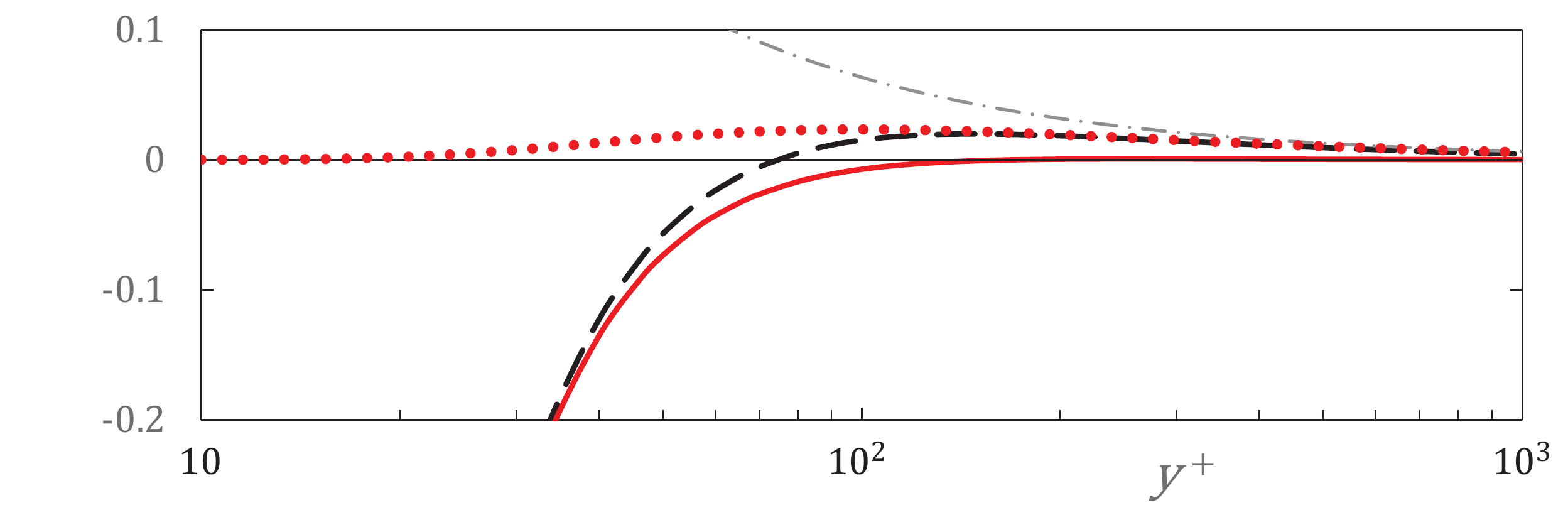}
\caption{(color online)  $\hat{U}^+_{\mathrm{M}} - [\kappa_\mathrm{M}^{-1}\ln(y^+)+B_{\mathrm{M}}]$ for $\kappa_\mathrm{M}=0.396$ and $B_{\mathrm{M}}=4.717$ ($S=905.86$), with (\textcolor{red}{---}) and without (- - -) subtracting the corrective term in (\ref{mMusker}) (\textcolor{red}{$\cdot \cdot \cdot$}) from $\hat{U}^+_{\mathrm{M}}$. $-\cdot -\cdot$, asymptotic approach of uncorrected $\hat{U}^+_{\mathrm{M}}$ to the log-law.}
\label{Masym}
\end{figure}

The basic Musker profile (\ref{Musker}) provides a good approximation to actual near-wall profiles with a logarithmic region at large $y^+$, but it also has shortcomings.
One of them is the slow asymptotic approach to the log-law as $\hat{U}^+_{\mathrm{M}}\to \kappa_\mathrm{M}^{-1}y^+ +B_\mathrm{M}+\kappa_\mathrm{M}^{-2}(y^+)^{-1}+\mathcal{O}(y^+)^{-2}$. This defect is irrelevant when fitting experimental data, but is of concern for the higher order asymptotic expansion of section \ref{sec3}. It is easily corrected by cancelling the $(y^+)^{-1}$ deviation of $\hat{U}^+_{\mathrm{M}}$ from the log-law at large $y^+$, resulting in the modified Musker profile

\begin{equation}
\label{mMusker}
\hat{U}^+_{\mathrm{mM}}\left(y^+; \kappa_\mathrm{M}, B_\mathrm{M}\right) = \hat{U}^+_{\mathrm{M}}\left(y^+; \kappa_\mathrm{M}, B_\mathrm{M}\right) - (\kappa_\mathrm{M}^2 y^+)^{-1}\,\exp(-100/y^+)
\end{equation}
As shown in figure \ref{Masym}, the effect is to ensure a clean log-law for $y^+$ beyond approximately 150.

Another, more prominent defect of the basic Musker profile is that it is too low in a region around $y^+$ of 30. This
has first been described by \citet{NagibChauhan2008}, who added to the Musker profile the ``hump'' function
\begin{equation}
\label{Hump}
\hat{H}_{\mathrm{NC}}(y^+; h_1, h_2, h_3) = h_1\,\exp\left[- h_2\,\ln^2(y^+/h_3)\right]\quad ,
\end{equation}
with the original parameters $h_1=0.351$, $h_2=1$ and $h_3=30$. Their addition of a higher order term, behaving as $-0.5\,\beta (y^+)^2/\Reytau$ for $y^+\to 0$ with $\beta$ the pressure gradient parameter, is however not consistent, if the Musker profile is used as an approximation for the \textbf{leading order} of the inner asymptotic expansion of $U^+$ (see also section \ref{sec4}).

\section{\label{App2} Comment on the dimensional analysis of \cite{Luchini17}}

\citet{Luchini17} \citep[see also][]{Luchini18} has used dimensional analysis to derive a higher order, linear pressure gradient correction of the log-law. His analysis is briefly summarized here to pinpoint where it goes wrong.

Luchini's application of the Buckingham $\Pi$ theorem starts with the five dimensional variables $\{\breve{U}_y , \breve{y}, \breve{u}_\tau, \breve{p}_x, \breve{\rho} \}$. One immediately notes that this starting list of variables implicitly contains the hydraulic diameter $\breve{D}_H \equiv - 4 \breve{\tau}_w/\breve{p}_x$, while Luchini explicitly excluded the other outer length scale $\breve{L}$ (e.g. the channel half-height or pipe radius). From this starting list, two non-dimensional $\Pi$'s are obtained ,
\begin{equation}
\Pi_1 = \frac{\breve{y}\,\breve{U}_y}{\breve{u}_{\tau}}~,\quad \Pi_2 = - \frac{\breve{p}_x\,\breve{y}}{\breve{\tau}_w} \equiv 4\,\frac{\breve{y}}{\breve{D}_H} \,\,,
\label{Lu1}
\end{equation}
which are related by the functional relation $\Pi_1 = F(\Pi_2)$. Assuming that the function $F$ is analytic around $\Pi_2 = 0$, Luchini Taylor-expanded $F$ around $\Pi_2 = 0$ and truncated the series after the linear term $\propto \Pi_2$, which leads to equation (6) in \cite{Luchini17} and, after non-dimensionalization and integration with respect to $y^+$, to
\begin{equation}
U^+ = \kappa^{-1}\,\ln{y^+} + B + 4 A_1 \frac{y^+}{D_H^+}~.
\label{Lu2}
\end{equation}
It is between equations (6) and (7) of \cite{Luchini17} that the analysis goes wrong, when he replaces the single outer length scale $D_H^+$ in the last term of (\ref{Lu2}) by $\beta\,(4\,L^+)^{-1}$ and implies that $\beta = 4\,L^+/D_H^+$ and $L^+$ can be chosen independently, in contradiction with the exclusion of $\breve{L}$ from the list of starting variables for the Buckingham's $\Pi$ theorem. In other words, equation (7) of \cite{Luchini17} is only valid for a \textbf{fixed} ratio of $D_H^+$ and $L^+ \equiv \Reytau$, i.e. a fixed $\beta$.

Introducing also $\breve{L}$ in the starting list of variables for the application of Buckingham's $\Pi$ theorem adds a third $\Pi_3 = \beta$ to the list of the parameters (\ref{Lu1}). As a consequence, nothing prevents the three parameters $\kappa$, $B$ and $A_1$ in equation (\ref{Lu2}) from becoming non-universal functions of $\beta$, as already suggested for $\kappa$ and $B$ by \cite{NagibChauhan2008}, \cite{Monk17} and \cite{Monkarxiv19}, for instance.

\bibliographystyle{jfm}
\bibliography{Karmanaps}

\end{document}